\documentclass[12pt]{article}
\usepackage[a4paper, margin=1in]{geometry}
\usepackage{multicol}%
\usepackage{multirow}%
\usepackage{url}
\usepackage{calc}%
\usepackage{xcolor}%
\usepackage{float}%

\usepackage{caption}

\usepackage[numbers,authoryear]{natbib}
\usepackage{amsmath,amsthm,boites,amssymb,amsbsy}
\usepackage{booktabs}
\usepackage{etoolbox}
\usepackage{subcaption}
\usepackage{authblk}
\patchcmd{\somecommand}{\vert}{\originalvert}{}{}
\usepackage{mathptmx} 

\PassOptionsToPackage{colorlinks=true, linkcolor=blue, urlcolor=blue, citecolor=blue}{hyperref}
\usepackage{hyperref}
\hypersetup{bookmarksopen=true,%
              bookmarksdepth=3}%

\usepackage[final]{listings}
\lstdefinestyle{C}{
   xleftmargin = 1.25cm,
   basicstyle  = \small\ttfamily,
   breaklines  = true,
   columns     = fullflexible,
   texcl       = true
}

\AtBeginDocument{\counterwithin{lstlisting}{section}}
\usepackage[acronym]{glossaries}
\makeglossaries

\usepackage{float}
\usepackage{algpseudocode}
\usepackage{algorithm}
\usepackage{fix-cm}

\usepackage{cleveref}
\crefname{lstlisting}{Listing}{Listings}
\Crefname{lstlisting}{Listing}{Listings} 
\usepackage{longtable}
\usepackage{colortbl}
\definecolor{codegreen}{rgb}{0,0.6,0}
\definecolor{codegray}{rgb}{0.5,0.5,0.5}
\definecolor{codepurple}{rgb}{0.58,0,0.82}
\definecolor{light-gray}{HTML}{F0F0F0}

\definecolor{backcolour}{rgb}{0.95,0.95,0.92}
\makeatletter
\newcolumntype{K}[1]{>{\raggedright\arraybackslash}p{#1}}
\makeatother
\usepackage{tikz}

\usetikzlibrary{patterns}
\usetikzlibrary{decorations.pathreplacing, decorations.markings}
\usepackage{pgfplots}
\pgfplotsset{compat=1.17}
\usepackage{tcolorbox} 

\usepackage{array}
\usepackage{placeins}

\counterwithin*{footnote}{section}
\renewcommand{\thefootnote}{\fnsymbol{footnote}}

\makeatletter
\newcommand\code{\bgroup\@makeother\_\@makeother\~\@makeother\$\@codex}

\def\@codex#1{{\normalfont\ttfamily\hyphenchar\font=-1 #1}\egroup}
\let\proglang=\textsf
\makeatother

\usepackage{algorithm, algorithmicx, algpseudocode}

\def\@jissue{}%
\def\jissue#1{\gdef\@jissue{#1}}%

\def\@licenseVal{}%
\def\licenseVal#1{\gdef\@licenseVal{#1}}%

\def\@supplinks{}%
\def\supplinks#1{\gdef\@supplinks{#1}}%

\def\@coninterest{}%
\def\coninterest#1{\gdef\@coninterest{#1}}%

\def\@confinfo{}%
\def\confinfo#1{\def\@confinfo{#1}}%

\def\@cdate{}%
\def\cdate#1{\gdef\@cdate{#1}}%

\def\@cname{}%
\def\cname#1{\gdef\@cname{#1}}%

\def\@cloc{}%
\def\cloc#1{\gdef\@cloc{#1}}%

\newcommand{\xleftrightarrow}[2][]{\ext@arrow 3399\leftrightarrowfill@{#1}{#2}}%

\def\emph#1{\textit{#1}}

\makeatletter
\def\@codex#1{{\normalfont\ttfamily\hyphenchar\font=-1 #1}\egroup}
\let\proglang=\textsf

\newcommand{\pkg}[1]{{\fontseries{m}\fontseries{b}\selectfont #1}}
\makeatother

\algnewcommand\algorithmicforeach{\textbf{for each}}
\algdef{S}[FOR]{ForEach}[1]{\algorithmicforeach\ #1\ \algorithmicdo}

\algnewcommand{\Inputs}[1]{%
  \State \textbf{Inputs:}
  \Statex \hspace*{\algorithmicindent}\parbox[t]{.8\linewidth}{\raggedright #1}
}
\algnewcommand{\Data}[1]{%
  \State \textbf{Data:}
  \Statex \hspace*{\algorithmicindent}\parbox[t]{.8\linewidth}{\raggedright #1}
}
\algnewcommand{\Outputs}[1]{%
  \State \textbf{Outputs:}
  \Statex \hspace*{\algorithmicindent}\parbox[t]{.8\linewidth}{\raggedright #1}
}
\algnewcommand{\Initialize}[1]{%
  \State \textbf{Initialize:}
  \Statex \hspace*{\algorithmicindent}\parbox[t]{.8\linewidth}{\raggedright #1}
}
\algtext*{EndWhile}
\algtext*{EndFor}
\algtext*{EndIf}

\usepackage[acronym]{glossaries}
\setacronymstyle{long-sc-short} 
\newacronym{FREQ}{FREQ}{number of accidents}
\newacronym{SVM}{SVM}{support vector machine}
\newacronym{LENGTH}{LENGTH}{segment length in miles}
\newacronym{FC}{FC}{funcional class (1 = local, 2 = collector, 3 = arterial, 4 = principal arterial, 5 = interstate)}
\newacronym{ROUTE}{ROUTE}{route number}
\newacronym{ACCESS}{ACCESS}{segment access control (0 = none, 1= partial, 3 = full)}
\newacronym{FRICTION}{FRICTION}{friction value (0 to 100 with 100 being high)}
\newacronym{AADT}{AADT}{annual average daily travel}
\newacronym{LNAADT}{LNAADT}{logarithm of annual average daily travel}
\newacronym{L50}{L50}{level terrain, longer than 50\% of the corresponding road length}
\newacronym{SW}{SW}{shoulder width}
\newacronym{ADTLANE}{ADTLANE}{average daily travel per lane}
\newacronym{PEAKHR}{PEAKHR}{the percentage of daily transit in peak hour}
\newacronym{INTECHAG}{INTECHAG}{number of interchanges in the segment}
\newacronym{GRADEBR}{GRADEBR}{number of grade breaks in the segment}
\newacronym{CURVES}{CURVES}{number of curves in the segment}
\newacronym{CURVE P}{CURVE P}{curvature presence}
\newacronym{SINGLE}{SINGLE}{daily percentage of single unit trucks}
\newacronym{MEDWIDTH}{MEDWIDTH}{median width}
\newacronym{MPB}{MPB}{mean prediction bias}
\newacronym{Lter}{Lter}{level terrain}
\newacronym{Rollter}{Rollter}{Terrain type is rolling}
\newacronym{VS Curve}{VS Curve}{very sharp curvature}
\newacronym{S Curve}{VS Curve}{sharp curvature}
\newacronym{SEG L}{SEG L}{segment length}
\newacronym{Mtter}{Mtter}{Terrain type is mountainous}
\newacronym{URB}{URB}{indication of an urban road}
\newacronym{MXMEDSH}{MXMEDSH}{maximum median shoulder width in feet}
\newacronym{MIMEDSH}{MIMEDSH}{minimum median shoulder width in feet}
\newacronym{MXGRADE}{MXGRADE}{maximum grade in the segment}
\newacronym{MIGRADE}{MIGRADE}{minimum grade in the segment}
\newacronym{MXGRDIFF}{MXGRDIFF}{maximum grade difference in the segment
}
\newacronym{PM}{PM}{median is present}
\newacronym{GBRPM}{GBRPM}{number of grade breaks per mile}
\newacronym{DOUBLE}{DOUBLE}{daily percentage of truck and trailer trucks}
\newacronym{LNADTMKM}{LNADTMKM}{logarithm of annual average daily travel by length in Km}
\newacronym{RS}{RS}{rural single carriageways}
\newacronym{RSMS}{RSMS}{rural single and medium speed}
\newacronym{RSHS}{RSHS}{rural single and high speed}
\newacronym{RSLS}{RSLS}{rural single and low speed}
\newacronym{FW}{FW}{formation width}
\newacronym{VS}{VS}{very sharp}
\newacronym{VS CURVE}{VS CURVE}{very sharp curvature}
\newacronym{NN}{NN}{neural network}
\newacronym{SPF}{SPF}{safety performance function}
\newacronym{HISNOW}{HISNOW}{segments where average snowfall per month is above 1 inch}
\newacronym{RT}{RT}{rolling terrain}
\newacronym{SP}{SP}{speed limit in Km/Hr}
\newacronym{MSP}{MSP}{indication of medium speed limit (>= 50 and < 100)}
\newacronym{LNLYR}{LNLYR}{logarithm of length by year}
\newacronym{LNKMYR}{LNKMYR}{logarithm of length in Km by year}
\newacronym{ATLM}{ATLM}{presence of Audio Tactile Lane Marking}
\newacronym{MCV}{MCV}{multi-combination vehicular traffic}
\newacronym{ArtVeh}{ArtVeh}{number of articulated vehicles}
\newacronym{FSI}{FSI}{fatal or serious injuries}
\newacronym{Nlanes}{Nlanes}{number of lanes}
\newacronym{NLANES}{NLANES}{number of lanes}
\newacronym{TANGENT}{TANGENT}{tangent length in the segment}
\newacronym{INCLANES}{INCLANES}{number of lanes in increasing milepost direction}
\newacronym{Length}{Length}{length of road in km}
\newacronym{LENYR}{LENYR}{length of road by period of analysis}
\newacronym{LENGZ}{LENGZ}{length of road in km}
\newacronym{ZI}{ZI}{Zero-Inflated}
\newacronym{ML}{ML}{machine learning}
\newacronym{rf}{rf}{random forrest}
\newacronym{GP}{GP}{Generalised Poisson}
\newacronym{MAE}{MAE}{mean absolute error}
\newacronym{LSP}{LSP}{Low Speed Highways}
\newacronym{COMP}{COMP}{Conway-Maxwell Poisson}
\newacronym{SA}{SA}{Simulated Annealing}
\newacronym{DE}{DE}{Differential Evolution}
\newacronym{HS}{HS}{Harmony Search}
\newacronym{MSL}{MSL}{maximum simulated log-likelihood}
\newacronym{MACML}{MACML}{maximum approximate composite marginal likelihood}
\newacronym{GA}{GA}{Genetic Algorithm}
\newacronym{PSO}{PSO}{Particle Swarm Optimisation}
\newacronym{BIC}{BIC}{Bayesian Information Criterion}
\newacronym{AIC}{AIC}{Akaike's Information Criterion}
\newacronym{DIC}{DIC}{Deviance Information Criterion}
\newacronym{PCA}{PCA}{Principal Component Analysis}
\newacronym{Pois}{Pois}{Poisson}
\newacronym{NB}{NB}{Negative Binomial}
\newacronym{LOCAL}{LOCAL}{local roads}
\newacronym{COLLECTOR}{COLLECTOR}{collector roads}
\newacronym{US}{US}{urban single carriageways}
\newacronym{UD}{UD}{urban double carriageways}
\newacronym{HSP}{HSP}{high-speed limit $\geq$ 100km/hr}
\newacronym{C50}{C50}{significant curvature (longer than 50\% of the segment)}
\newacronym{LW}{LW}{lane width}
\newacronym{AVEPRE}{AVEPRE}{average precipitation of rainfall}
\newacronym{MtTer}{MtTer}{mountainous terrain}
\newacronym{MtCurve}{MtCurve}{mountainous terrain with significant curvature}
\newacronym{CRP}{CRP}{correlated random parameters}
\newacronym{RPHM}{RPHM}{random parameters with heterogeneity in the means}
\newacronym{CRPHM}{CRPHM}{correlated random parameters with heterogeneity in the means}
\newacronym{RP}{RP}{random parameters}
\newacronym{RMSE}{RMSE}{root-mean squared error}
\newacronym{WCP}{WCP}{wide centerline greater than 50\% of road length} 
\newacronym{SLOPE}{SLOPE}{Segment slope (0=flat, 1=slight, 2=medium, 3=high)}
\newacronym{MSPE}{MSPE}{mean square prediction error}
\newacronym{MSE}{MSE}{mean squared error}
\newacronym{MCRVRS}{MCRVRS}{\gls{C50} and \gls{RS}}
\newacronym{RD}{RD}{rural dual carriageway}
\newacronym{RF}{RF}{Random Forest}
\newacronym{GRP}{GRP}{grouped random parameters}
\newacronym{SMGRP}{SMGRP}{single means grouped random parameters}
\newacronym{MMGRP}{MMGRP}{multiple means grouped random parameters}
\newacronym{WCLTs}{WCLTs}{wide centre line treatments}
\newacronym{FC1}{FC1}{local road type}
\newacronym{FC2}{FC2}{collector road type}
\newacronym{MP}{MP}{mathematical program}
\newacronym{GOF}{GOF}{goodness-of-fit}
\newacronym{M}{M}{Roadtype is motorway}
\newacronym{DX32}{DX32}{days with temperature less than 32 F}
\newacronym{PRCP}{PRCP}{total monthly precipitation in inches}
\newacronym{M PRCP}{M PRCP}{maximum monthly precipitation}
\newacronym{FW RS}{FW RS}{formation width on rural single carriageways}
\newacronym{DSND}{DSND}{Days with snowfall}
\newacronym{DSNW}{DSNW}{days with snowfall greater than 1 inch}
\newacronym{DP10}{DP10}{days with precipitation greater than 1 inch}
\newacronym{DP01}{DP01}{days with precipitation greater than 0.1 inch}
\newacronym{HSM}{HSM}{Highway-Safety Manual}
\newacronym{RUMBLE}{RUMBLE}{presence of rumble strip}
\newacronym{LEN YR}{LEN YR}{segment length multiplied by period of analysis}
\newacronym{Curve50}{Curve50}{significant curvature (moderate to very sharp which is longer than 50\% of the segment)}
\newacronym{LENKM}{LENKM}{length in km}
\newacronym{CPM}{CPM}{curvature per mile}
\newacronym{HQIC}{HQIC}{Hanan-Quinn Criterion}
\newacronym{LOWPRE}{LOWPE}{average precipitation in a month is less than 1.5 inches}
\newacronym{INTPM}{INTPM}{interchanges per mile}
\newacronym{SPEA2}{SPEA2}{Strength Based Pareto Evolutionary Algorithm 2}
\newacronym{COA}{COA}{Chaos Optimisation Algorithm}
\newacronym{COM-Poisson}{COM-Poisson}{Conway-Maxwell-Poisson}
\newacronym{MCMC}{MCMC}{Markov Chain Monte Carlo}
\newacronym{SPEED}{SPEED}{speed in mp/hr}

\newacronym{HM}{HM}{harmony memory}
\newacronym{PAP}{PAP}{pitch-adjustment proportion}
\newacronym{PAI}{PAI}{pitch-adjustment index}
\newacronym{PAR}{PAR}{pitch-adjustment rate}
\newacronym{HMCR}{HMCR}{harmony memory consideration rate}
\newacronym{HMS}{HMS}{harmony memory size}
\newacronym{GENP}{GENP}{genetic program}
\newacronym{CR}{CR}{crossover rate}
\newacronym{PS}{PS}{population size}
\newacronym{NOV}{NOV}{month of November}
\newacronym{P SHLDER}{P SHLDER}{paved shoulder}
\newacronym{HSLOPE}{HSLOPE}{high levels of slopes}
\newacronym{M MEDWIDTH}{M MEDWITH}{average levels of medwidth}
\newacronym{MX PRCP}{MX PRCP}{maximum precipitation encountered from the month (inches)}
\newacronym{GLM}{GLM}{generalized linear model}
\lstdefinestyle{mystyle}{
    backgroundcolor=\color{backcolour},   
    commentstyle=\color{codegreen},
    keywordstyle=\color{magenta},
    numberstyle=\tiny\color{codegray},
    stringstyle=\color{codepurple},
    basicstyle=\footnotesize,
    breakatwhitespace=false,         
    breaklines=true,                 
    keepspaces=true,                 
    numbers=left,                    
    numbersep=5pt,                  
    showspaces=false,                
    showstringspaces=false,
    showtabs=false,                  
    tabsize=2
}

\begin{document}
\setcounter{lstlisting}{0}
\title{MetaCountRegressor: A python package for extensive analysis and assisted estimation of count data models}

\author[1]{Zeke Ahern}

\author[2]{Paul Corry}

\author[1]{Alexander Paz}

\title{Assisted and Extensive Estimation of Data Count Models}

\affil[1]{Civil and Environmental Engineering, Queensland University of Technology, Queensland, Australia}

\affil[2]{Mathematical Sciences, Queensland University of Technology, Queensland, Australia}





\abstract{Analyzing and modeling rare events in count data presents significant challenges due to the scarcity of observations and the complexity of underlying processes, which are often overlooked by analysts due to limitations in time, resources, knowledge, and the influence of biases. This paper introduces \pkg{MetaCountRegressor}, a  Python package designed to facilitate predictive count modeling of rare events guided by metaheuristics. The \pkg{MetaCountRegressor} package offers a wide range of functionalities specifically tailored for the unique characteristics of rare event prediction. This package offers a collection of metaheuristic algorithms that efficiently explore the solution space, facilitating effective optimisation and parameter tuning. These algorithms are specifically engineered to deal with the inherent challenges of modeling rare events for predictive purposes, and capturing causative effects that are easily interpretable. State-of-the-art models are produced by the decision-based optimization framework. This includes the ability to capture unobserved heterogeneity through random parameters and allows for correlated and grouped random parameters. It also supports a range of distributions for the random parameters, and can capture heterogeneity in the means. The package also supports panel data, among other features, and serves as a systematic framework for analysts to discover optimization-driven results, saving time, reducing biases, and minimizing the need for extensive prior knowledge.}

\maketitle

\renewcommand\thefootnote{}

\renewcommand\thefootnote{\fnsymbol{footnote}}
\setcounter{footnote}{1}
\onecolumn
\section{Introduction: Resources for Fitting Models}\label{sec1}
Predictive count models are used in various fields, including but not limited to insurance, retail, healthcare, road safety, and transportation\citep{Sawtelle2023DriverMaine}. For example, count models can be used to predict the number of admissions to hospitals for individuals with eating disorders \citep{Kim2023UsingDisorders}, estimate the number of penalties scored in a soccer season \citep{Sadeghkhani2022K1K2InflatedMatches}, and analyze fire frequency \citep{Marchal2017ExploitingCanada}, among other important real-world applications. The primary objective of predictive count models is learn about patterns, capture trends, and/or understand behavior using data. Through the analysis of historical counts and consideration of various potential contributory factors, these models uncover relationships and factors that influence count outcomes. Their development aims to provide accurate predictions, facilitate proactive decision-making, optimize resource allocation, and enhance overall operational efficiency across industries \citep{Mannering2014AnalyticDirections}.

Open-source software provides a wide array of statistical packages encompassing various features, including hurdle implementation \citep{Zeileis2008RegressionR}, underdispersion modeling \citep{Sellers2010AData}, and mixed effects capabilities \citep{Faraway2016ExtendingEdition}. In addition to the aforementioned software packages, there are numerous other notable options worth mentioning. One such example is the \pkg{gamlss} \citep{Stasinopoulos2008GeneralizedR}, which is valuable when analyzing data exhibiting non-linearities, heteroscedasticity, and non-normal error distributions. Another noteworthy option is \pkg{VGAM} \citep{Yee2010TheAnalysis}, which specialises in the analysis of multivariate data.  However, these software packages have a significant limitation. They do not fully capture unobserved heterogeneity, which requires consideration of more complex modeling aspects such as random parameters and heterogeneity in the means. Furthermore, their exists little assistance provided by these packages in developing such complex models such as local-search methods used in basic regression packages like \pkg{leaps} \citep{Miller2002SubsetRegression}. For instance, the \pkg{MASS} package is capable of performing step-wise model selection, utilizing a straightforward heuristic to enhance the \gls{AIC} metric. It sequentially adds or removes contributing factors within the model formula \citep{Venables1999ModernS-PLUS}. 

Although random parameters for count data are available in \pkg{Nlogit} packages, this feature is not open-source \citep{Hilbe2006A4.0}, and the distributional assumptions for these random parameters are quite limited, offering only Uniform, Normal and its variations. Other hierarchical software is available, though its application is often specific to particular use cases. For example, the \pkg{unmarked} package in R provides a versatile modeling framework for a range of survey methods, including site occupancy sampling, repeated counts, distance sampling, removal sampling, and double observer sampling. Furthermore, is proficient at model selection by utilizing a local search method, as demonstrated by \cite{Fiske2011Unmarked:Abundance}. Model selection, especially in hierarchical models, can be computationally intensive as it often necessitates testing many models and selecting the best without a systematic or formal approach. To address this, the \pkg{SMS} (Smart Model Selection) software package was developed. \pkg{SMS} significantly reduces computational calculations and runtime for model selection in phylogenetic analysis by employing a heuristic strategy through model filtering \citep{Darriba2012JModelTestComputing}. It achieves efficiency without significantly compromising the results compared to exhaustive calculations \citep{Lefort2017SMS:PhyML}. More relevant to the proposal in this paper, \pkg{glmulti} was designed to automate model selection and multi-model inference with \glspl{GLM}. It offers functionalities such as exhaustive screening and a compiled genetic algorithm method to escape local optima. Such capabilities are not found in model selection software that use heuristic implementations \citep{Calcagno2010Glmulti:Models}.

Despite these developments, an analyst still needs to create and test specifications using problem context knowledge, experience, and statistics \citep{Paz2019SpecificationFramework}. For example, even in a \gls{GLM}, discovering the best set of contributing factors to include in the model can be a complex task \citep{Paz2019SpecificationFramework}. Some features may cause overfitting or be non-significant, and it is challenging to discover the final specification \citep{Lord2010TheAlternatives}. 

Typically, an analyst might introduce their own biases into the model based on knowledge and experience \citep{Behara2021AQueensland}.Although knowledge is indeed beneficial, this approach can potentially neglect other hypotheses in the process, since analysts typically employ a naive local search method \citep{Mannering2020BigAnalysis}. An approach to overcome local search methods is to wrap the regression problem around an evolution algorithm \citep{Oliveira2010GA-basedEstimation,Akbar2024GeneticProjects}.

For example, a study compared the effectiveness of greedy, metaheuristic, and evolutionary search algorithms, for regression test case prioritisation across six problems of varying sizes \citep{Li2007SearchPrioritization}. The findings revealed that \glspl{GA} outperformed greedy-based approaches despite the presence of non-linearities in the decision-making process within the regression testing search space. This superiority can be attributed to the inherent global-search capabilities of \glspl{GA}, which effectively can escape sub-optimal regions \citep{Deb2002ANSGA-II}. The efforts described by \cite{Li2007SearchPrioritization}, although effective in determining an efficient search strategy, do not encompass the broader scope of count prediction estimation. 
Additionally, these efforts do not address the diverse range of state-of-the-art modeling and estimation capabilities required for developing heterogeneous models \citep{Mannering2020BigAnalysis,Ali2022AEnvironment,Behara2021AQueensland}. 

In particular, a groundbreaking approach in the field of heterogeneous modeling is the use of search-based implementation, exemplified by the innovative package \pkg{searchlogit} \citep{Beeramoole2023ExtensiveModels}. This software provides an assisted metaheuristic implementation for incorporating the latest modeling practices for Mixed Logit models. By leveraging the power of search-based techniques, \pkg{searchlogit} offers valuable assistance in addressing complexities associated with state-of-the-art modeling in heterogeneous contexts. While the availability of software like \pkg{searchlogit} is highly valuable for Mixed Logit models, there remains a pressing need for specialised software that caters specifically to count prediction analyzes.

The \pkg{MetaCountRegressor} package has been developed and tested in \cite{Ahern2024ExtensiveModels}, where extensive hypothesis testing was conducted to propose count models for road safety crash frequency that are capable of identifying hierarchical models with random parameters. Since then, the software has undergone various enhancements and functional capabilities, explained throughout this paper. While the work outlined in this paper focuses on fitting predictive count models, the \pkg{MetaCountRegressor} package can also be applied to linear regression. An implementation of \pkg{MetaCountRegressor} for linear regression can be found in \cite{FlorianHeraud2025UncoveringBehavior}, which focuses on identifying the best specification for differences in parking payments, as well as the relationship between the amount spent and the remaining time.


\section{Functional Capabilities to Address Unobserved Hetrogeneity}\label{sec2}

\section{Models and software} \label{secpap4:models}
The models developed in \pkg{MetaCountRegressor}  are based on the fundamental assumptions of the accessible closed-form models used for count prediction, which include, \gls{Pois} and \gls{NB}. Therefore, consider the set of models $\mathcal{M}$ as $\mathcal{M} = \{Pois, NB\}$  Thus, it is assumed that the count response variable, denoted as $y_i$, can be predicted by a conditional modeling equation. Specifically, given an appropriate distribution $m$ from the set $\mathcal{M}$,  $y_i$ is approximated as $\lambda_i$, where $\lambda_i$ is modeled by \Cref{eq4:pap4simple_fixed}

 \subsection{Hierarchical Count Data Formulation and Capabilities}
Consider the following representation of a simple fixed-effects model for the expected number of crashes, denoted as $\lambda_i$, observed on the $i$-th road segment, formulated by a \gls{NB} closed-form model:
\begin{align}
\lambda_{i|m} = \exp\left[\beta^F{X_i}^F +\varsigma_i\right]  & &\forall i \in \mathcal{I}, m' \in \mathcal{M}^E: \kappa_m = 1, m \in \mathcal{M} \label{eq4:pap4simple_fixed}
\end{align}
 In this model, ${X_i}^F$ represents a vector that encompasses the likely contributing factors associated with road segment $i$. The dimension of this vector is determined by the number of likely contributing factors in the set $\mathcal{K}_F$, denoted as $|\mathcal{K}_F|$. The elements within $\mathcal{K}_F$ are the relevant factors that influence crash occurrences, but selection of them are decision variables.  As a result, in the mathematical programming formulation, the selection process is treated by considering the contributing factors as decision variables, thus forming the appropriate $\mathcal{K}_F$ from the set of all possible potential contributing factors $\mathcal{K}$.  Unlike many modeling processes that completely overlook variable selection \citep{Hunter2023ScienceCare}, this software handles it as an integral part of the process.

In addition to the likely contributing factors included in the fixed effects model, the framework incorporates an error term denoted by $\varsigma_i$. This error term follows a distribution that is commonly assumed to be gamma distributed if the analyst's suggests that model $m$  from $\mathcal{M}$ is of \gls{NB} type \citep{Lord2005PoissonTheory}. In this case, the distribution of the error term is conditioned by the selection of $m$ from $\mathcal{M}^E$, which represents a set of error assumed terms stemming from the assumptions of a closed-form model. The error term in the model represents the inherent variability in the observed counts. However, it does not account for potential unobserved effects among the likely contributing factors. To address this limitation, many studies have utilized a traditional non-correlated \gls{RP} \footnote{Our paper \cite{Ahern2024ExtensiveModels} deals with extensive hypothesis testing for RP models} or a more complex \gls{CRP} approach, which allows correlations to exist between random parameters \citep{Huo2020AAnalysis, Hou2018AnalyzingApproach}\footnote{Our paper \cite{Ahern2024Multi-objectiveModels} deals with correlated random parameters}. Despite the increased complexity of the \gls{CRP} model, the same likely contributing factors are used in both the \gls{RP} and \gls{CRP} approach. To incorporate these considerations, the simple fixed effects model presented in \Cref{eq4:pap4simple_fixed} is extended for random parameters in \Cref{eq4:pap41}.
\begin{align}
\lambda_{i|\omega_i, m} = \exp\left[\beta^F{X_i}^F +{\beta_i}^R {X_i}^R+\varsigma_i\right]  & &\forall i \in \mathcal{I}, m' \in \mathcal{M}^E: \kappa_m = 1, m \in \mathcal{M} \label{eq4:pap41}
\end{align}
To specifically focus on the structure of the \gls{RP} model, the attention needs to be directed towards ${\beta_i}^R$ and its associated counterparts. For example, in \Cref{eq4:pap41}, the model is conditioned by a vector of random parameters $\omega_i$, which has a size of $|\mathcal{K}_P|\times 1$. These random parameters allow observations to be varied based on a pre-specified distribution. In this formulation, they can be unique and different for each contributory factor from a set of distributions $\mathcal{D}$. This information only captures the number of random parameters, and the \gls{RP} structure in \Cref{eq4:pap41} can be manipulated to form the \gls{CRP} models, along with \gls{HM}, enabling the existence of additional heterogeneity in the means of the random parameters. To adequately account for these correlations and heterogeneity in means, an appropriate formulation of ${\beta_i}^R$ is required \citep{Saeed2019AnalyzingHighways} and presented in \Cref{eq4:pap4cor_ran} \footnote{Our paper \cite{Ahern2024ExtensiveModels} deals with heterogeneity in the means and grouped random parameters}.
\begin{align}
\beta_{i, k}^R =&\beta_k^{R}+ \Delta_k Z_{i,k, n} +C_k\omega_{i,k} & \forall i \in \mathcal{I},\forall k \in \mathcal{K}_P, n \in \mathcal{Z}_k \label{eq4:pap4cor_ran}
\end{align}
Wherein, the term $Z_{i, k}$ (consisting of $n$ elements denoted as $Z_{i, k, n}$) represents a vector of likely contributing factors that encompass heterogeneity in the mean value of $\beta_{i, k}^R$ by incorporating $n$ additional factors for all $n \in \mathcal{Z}_k$.  Here, $\mathcal{Z}_k$ denotes a set of sets with a length of $|\mathcal{K}_P|$ that captures the likely contributing factors associated with the heterogeneity of means at element $k$. Therefore, the inclusion of $n \in \mathcal{Z}_k$ allows for the consideration of the $Z_{i,k,n}$ variable, which provides supplementary information beyond what is already captured by the variables $X_{i}^R$. Furthermore, $\Delta_k$ is a vector of estimable parameters that captures heterogeneity in the means. The size of $\Delta_k$ corresponds to the number of likely contributing factors in $\mathcal{Z}_k$. The correlations between the random parameters are specifically accounted for through the variance-covariance matrix $C$, structured by:
\begin{align}
C = \left[ \begin{matrix} (\sigma_{1,1})^2 &  &  & \ &  \\ \sigma_{2,1}^2 & (\sigma_{2,2})^2 &  &  &  \\ \vdots & \vdots & \ddots &  &  \\ \sigma_{k-1,1}^2 & \sigma_{k-1,2}^2 & \cdots & (\sigma_{k-1,p-1})^2 &  \\ \sigma_{i,1}^2 & \sigma_{i,2}^2 & \cdots & \sigma_{k,p-1}^2 & (\sigma_{k,p})^2 \end{matrix} \right] & &\forall k, p \in \mathcal{K}_P\label{eq:pap4C:matrix}
\end{align} 
Here, $C$ is the variance-covariance matrix of the random parameters vector $\beta_i^R$. It is a square matrix of size $|\mathcal{K}_P| \times |\mathcal{K}_P|$. The diagonal elements of $C$ represent the variance of each individual random parameter, while the off-diagonal elements represent the covariance between each pair of random parameters. Note, \Cref{eq4:pap4cor_ran} is evaluated for all values of $k$ in the set $\mathcal{K}_P$. Consequently, for the $k^{th}$ random parameter, the coefficients $\beta_{i,k}$ vary according to the following equation:
\begin{align}
   \beta_{i, k}= \beta_k + \Delta_kZ_{i, k, n}   + \sum_{p=1}^{\mathcal{K}_P} \sigma_{k,p} \omega_{i, k} &&  i \in \mathcal{I}, k \in \mathcal{K}, \forall n \in \mathcal{Z}_k
\end{align}
Where $n$ represents the number of heterogeneous terms in the means for each $k$ random parameters. 

In addition, another decision is to consider whether to utilize grouped random parameters or traditional random parameters. This decision involves determining the presence of parameters at the group level, represented as $\hat{\omega}_g$, where $\hat{\omega}_g$ is a vector of groups denoted as $\mathcal{G}$. The groupings are determined by $\eta_i$, where $\omega_{i,k}$ represents a vector of random parameters. This can be used to indicate the groups to which each observation belongs. When $\omega_{i,k}$ is interpreted as a grouped random parameter, it signifies that
\begin{align}
    \omega_{i, k} = \hat{\omega}_{g, k} \quad &\forall i \in \mathcal{I}, \forall k \in \mathcal{K}_P: \varrho_k,  \forall g \in \mathcal{G} : g = \eta_i \label{eq4:grpap4} 
\end{align}
In the absence of explicitly stated groups, \cref{eq4:grpap4} simplifies to a traditional random parameters model at the observation level.  When considering grouped random parameters, decisions regarding the applicability of grouped random parameters for specific $k$ in the set of indices $\mathcal{K}_P$ are distinguished. To make these decisions, a vector of indicator variables, denoted as $\varrho$, is utilized. The variable $\varrho_k$ indicates the usage of random sample draws at the grouped level when true; otherwise, standard random sample draws are used at the observation level. 

In this decision-variable focused framework, a fully correlated random parameters model with heterogeneity in the means and grouped random parameters represents the highest hierarchical structure that can be estimated. However, it is also possible to test reduced specifications through adequate construction by the decision variables. Therefore, a mathematical programming problem is formulated to construct the most efficient modeling specifications, to facilitate the use of the \gls{CRPHM} with the \gls{GRP} model, if aligned with the objective function, and to allow specifications of lower-level hierarchies.





\subsection{Solution Algorithms Fundamentals}

\begin{figure}
    \centering
    \resizebox{0.5\textwidth}{!}{
    \begin{tikzpicture}
\tiny
  \foreach \x/\bit/\color in {1/${\alpha}_1 \cup r_1$/pink, 2/$\dots$/pink, 3/${\alpha}_K \cup r_K$/pink, 4/$\tau_1$/green!50, 5/$\dots$/green!50, 6/$\tau_K$/green!50, 7/$f_1$/orange!50, 8/$\dots$/orange!50, 9/$f_K$/orange!50, 10/$\kappa$/purple!20} {
    \draw[fill=\color] (\x-0.5, -0.5) rectangle (\x+0.5, 0.5);
    \node at (\x, 0) {\bit};
  }

 \draw[decorate,decoration={brace,amplitude=5pt,mirror}]
    (1-0.5, -0.5) -- node[below=6pt, text width=3cm, align=center] {Selected: Fixed/ Random Parameters / Correlated Random / Heterogeneity Effects or No Effect} (3+0.5, -0.5);
  \draw[decorate,decoration={brace,amplitude=5pt,mirror}]
    (4-0.5, -0.5) -- node[below=6pt, text width=3cm, align=center] {Selected Transformations} (6+0.5, -0.5);
  \draw[decorate,decoration={brace,amplitude=5pt,mirror}]
    (7-0.5, -0.5) -- node[below=6pt, text width=3cm, align=center] {Random Parameters Distributions} (9+0.5, -0.5);
  \draw[decorate,decoration={brace,amplitude=5pt,mirror}]
    (10-0.5, -0.5) -- node[below=6pt, text width=1.5cm, align=center] {Closed-form Model} (10+0.5, -0.5);

  \draw[decorate,decoration={brace,amplitude=5pt}]
    (1-0.5, 0.5) -- node[above=6pt, text width=3cm, align=center] {Choice from a set of discrete decisions} (3+0.5, 0.5);
  \draw[decorate,decoration={brace,amplitude=5pt}]
    (4-0.5, 0.5) -- node[above=6pt, text width=3cm, align=center] {Select from $T$} (6+0.5, 0.5);
  \draw[decorate,decoration={brace,amplitude=5pt}]
    (7-0.5, 0.5) -- node[above=6pt, text width=3cm, align=center] {Select From $D$} (9+0.5, 0.5);
  \draw[decorate,decoration={brace,amplitude=5pt}]
    (10-0.5, 0.5) -- node[above=6pt, text width=1.5cm, align=center] {Select From $M$} (10+0.5, 0.5);

\end{tikzpicture}}
\caption{A chromosome representation for the decision variables. The length is dependent on the number of potential contributing factors (K) that are available from the data.}
\label{fig:chromosome4}

\end{figure}
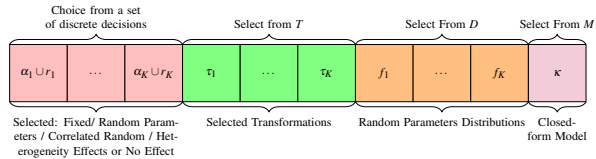

Discrete variants of three metaheuristics, namely \gls{DE}, \gls{SA}, and \gls{HS}, have been developed, implemented, and tested. In \pkg{MetaCountRegressor}, a chromosome representation\footnote{detailed in \cref{fig:chromosome4}} capturing the discrete characteristics of the problem, chromosome corresponds to a unique hypothesis classified by the mathematical program. For a more comprehensive understanding of this representation and other metaheuristic solution algorithms designed to tackle intricate multi-combinatorial optimisation problems, please refer to \cite{Abdel-Basset2018MetaheuristicReview}. Within this discrete framework, certain modifications were made to these algorithms. \Gls{SA}, being discrete by design, required minimal adjustments. However, discrete \gls{DE} and \gls{HS} were inspired by the works of \cite{Wu2018EnsembleVariants} and \cite{Kattan2010HarmonyNetworks} respectively, in order to adapt them to the proposed discrete problem setting. The logic and workflow of the algorithms available in this software can be found in \cref{alg:exppap4}.

The chromosome structure is illustrated in \cref{fig:chromosome4}.

\FloatBarrier

\section{Testing Example} \label{secpap4:illustrations}

For a tutorial of \pkg{MetaCountRegressor}, is provided in the PyPi storage \citep{Ahern2023MetacountregressorPyPI}(\href{https://pypi.org/project/metacountregressor/}{PyPi: Metacountregressor} ), which includes a detailed markdown script of implementations, parameters, settings and details to follow to enable a seamless experience. A link to the tutorial is also provided, see( \href{https://anonymous.4open.science/r/data-8257/Step%20By%20Step_Synthetic.ipynb}{Tutorial} 

\subsection{Installation}
\pkg{MetaCountRegressor} requires Python 3.10 and can be installed by \cref{install:2}.

\begin{lstlisting}[language=Python, caption = Installation of the package, label={install:2}]
pip install metacountregressor
\end{lstlisting}

\FloatBarrier
The \pkg{MetaCountRegressor} package revolves around two main classes (solution and metaheuristics):
\begin{enumerate}
\item The \code{solution} class: This class covers all the statistical calculations required to test unobserved heterogeneity count models with various modeling structures.
\item The \code{metaheuristics} class: This class represents the solution algorithm, which can utilize metaheuristic optimization methods such as \gls{DE} (Differential Evolution), \gls{SA} (Simulated Annealing), or \gls{HS} (Harmony Search) to guide the search process.\footnote{The \code{harmony\_search}, \code{differential\_evolution}, and \code{simulated\_annealing} functions are used to implement these metaheuristic methods.}
\end{enumerate}
 The following commands call upon both classes. Only one of the metaheuristics needs to be chosen to optimize the ObjectiveFunction from \code{solution}.  Hence, importing only one of these \code{harmony\_search, differential\_evolution, simulated\_annealing} from \code{metaheuristics} is required. See \ref{listing:import}, which presents how to import one or all the metaheuristics.

\begin{lstlisting}[language={Python}, caption ={Importing}, label={listing:import}] 
from metacountregressor.solution import ObjectiveFunction
from metacountregressor.metaheuristics import (harmony_search,differential_evolution,simulated_annealing)         
\end{lstlisting}

\FloatBarrier
\subsection{Default Optimization Setup}

Three metaheuristic algorithms are available for optimizing the frequency regression model. In addition, selecting the data, defining the variable \texttt{Y}, and determining an offset term (if necessary) are the only additional steps required to begin the search. The metaheuristic algorithm will search for the best specifications, minimising both the \gls{BIC} and \gls{MSPE} simultaneously. By default, when no specific arguments are provided, these objectives will guide the solution. The following code in this subsection will optimize the specifications using all the attributes in the available data, discovering what contributing factors are important and how they should be modeled.

The dataset used in this example is available at the following URL: \href{https://anonymous.4open.science/r/data-8257/Ex-16-4.csv}{Washington data}. This dataset has been widely used in various crash studies, such as those by \cite{Shankar1996StatisticalFreeways} and \cite{Milton1998TheFrequencies}.

The \code{ObjectiveFunction} requires two arguments \code{X} and \code{y}, where \code{X} is our data of explanatory variables, defined as a \pkg{pandas} dataframe, and y is the response (measured as counts, or frequencies). More information can be supplied utilizing *args.

In this example, we will illustrate the basic out of the box optimization and routine without *args specified.  See \hyperref[note:input_data]{Note: No Arguments} for more details

\begin{tcolorbox}[colback=black!5!white, colframe=orange!75!black, title=Note: No Arguments,label=note:input_data]
When inputting the data, standard checks should be in place, handling null elements, prior, among dropping the relevant columns, never for consideration \footnote{Example, an ID column or highly correlated column}. Decisions should be made based on the analyst's judgement and preferences to drop these columns, to define the complete set of \code{X}. Furthermore, \texttt{X} should only be defined with contributing factors considered necessary by the user; including irrelevant factors will lead the optimization framework to test hypotheses including those specific terms. For example, if the data contains information about postcodes (area codes or code numbers), and these are not of interest to your analysis, they should be excluded. 
\end{tcolorbox}

If not excluded, irrelevant information within the data may eventually be filtered out through the optimization routine based on the objective-focused approach. However, providing as much relevant information as possible aligns with user preferences and helps to achieve better results by reducing the number of potential hypothesis tests. This allows more intuitive hypotheses to be tested efficiently.

Accordingly, load the inputs and define \code{X} and \code{y}. See\cref{lis:load}, for syntax on how this is performed.
\begin{lstlisting}[caption={Loading in data (example syntax).}, label={lis:load}, language={Python}] 
import numpy as np
import pandas as pd
df = pd.read_csv("https://anonymous.4open.science/r/data-8257/Ex-16-3.csv")
y = df['FREQ']  # Frequency of crashes
X = df.drop(columns = ['FREQ']) # setup X based on data
\end{lstlisting} 

\FloatBarrier
\begin{tcolorbox}[colback=black!5!white, colframe=orange!75!black, title=Note: No Offset,label=note:input_data_offset]
In \cref{lis:load}, the definition of an offset variable is absent, unless there happens to exist a named offset column in the data called 'Offset'. To ensure the offset is taken into consideration in the optimization routine,  the offset needs to be added as an explanatory column. Otherwise, no offset will be used. 
\end{tcolorbox}
Defining an offset term is supported by the framework and helps ensure the focus remains on the relationship between the explanatory variables and the response variable.

Furthermore, once the offset term is set, any related columns should be omitted to avoid correlation issues. See \cref{listing:offset} for an example.
\begin{lstlisting}[language={Python}, caption ={Defining an offset variable (example syntax).}, label ={listing:offset}]
X['Offset'] = X['LENGTH']*np.log1p(X['AADT']) #Modify Here for Desired Offset
X = df.drop(columns=['LENGTH', 'AADT', 'ID'])    
\end{lstlisting} 
\FloatBarrier
There are two types of arguments to consider:
\begin{enumerate}
    \item \textbf{objective arguments}:  Relate directly to problem being solved
    \item \textbf{metaheuristic arguments}: Guide the search process during optimization
\end{enumerate}
Metaheuristic arguments are optional and not required to initiate the process, other than calling the metaheuristic class itself. However, they can be adjusted to customize the search process. If there is uncertainty about these, they can be omitted entirely and simply call the desired metaheuristic without specifying its arguments.

On the other hand, objective arguments are more critical. These include key aspects such as the objective to be minimized, how the data is split, and other problem-specific settings. For an example of how to define these arguments, refer to the arguments provided in \cref{listing:arguments}.

Moreover, there is significant flexibility in how these arguments are structured. For instance, objective arguments can include details about various aspects of the problem, such as the validation split or assumptions regarding complexity (e.g., heterogeneity), which are explained further in \cref{tab:definedargs}.

\begin{lstlisting}[language=Python, caption = {Customizing the arguments for the solution and the metaheuristic search.}, label = {listing:arguments}]
arguments = {'test_percentage': 0.3, 'test_complexity': [1,2,3,5]}
arguments_hs = {'_par': 0.3, '_hms': 20}
arguments_sa = None #Note: Supply the relevant arguments, otherwise default arguments will be used
arguments_de = None
\end{lstlisting} 
\FloatBarrier

The solution, with its arguments, can be embedded into the metaheuristic solution algorithm. \Cref{listing:search} provides the syntax for implementing each of the available search algorithms. Note that these are all extensive searches; typically, only one solution algorithm is required. The choice of solution algorithm should be based on the analyst's preferences and available resources. \footnote{For an in-depth examination of these algorithms, refer to \cite{Ahern2024ExtensiveModels}, which evaluates the algorithms and discusses their performance. The paper highlights that, if the focus is on optimizing a single objective, simulated annealing is recommended. For multiple objectives, both harmony search and differential evolution are suitable.}

\begin{lstlisting}[language=Python, caption = {Defining the objective function for harmony search, differential evolution or simulated annealing}, label = {listing:search}] 
    initial_solution = None
    obj_fun = ObjectiveFunction(X, y, **arguments)
    #perform harmony search
    results_hs = harmony_search(obj_fun, initial_solution, **arguments_hs)
    #perform differential evolution
    results_de = differential_evolution(obj_fun, initial_solution, **arguments_de)
    #perform simulated annealing
    results_sa = simulated_annealing(obj_fun, initial_solution, **arguments_sa)
\end{lstlisting}
\FloatBarrier
Once the code in \cref{listing:search} is executed, the search process will begin. During each iteration, a new model is estimated, and if successfully estimated, it will be printed console log. A folder is also created where  the model is run from, and it will log the best results.

\begin{tcolorbox}[colback=black!5!white, colframe=orange!75!black, title=Note: Printing Arguments Offset,label=note:printing]
Printing can be customised to never output, with verbose =0, output for every solution with verbose = 1, or output for every new best/pareto efficient solution with verbose = 2.

\end{tcolorbox}

\footnote{argument to turn off printing}.

The initial solution from the search process is presented as evidence (see \cref{tab:exam}). However, this model represents only the first solution tested. With each iteration, new solutions are evaluated, progressively refining the outcomes. This iterative process progressively refines alternative specificifications, which will lead to a final proposed model that should be considered by the analyst.

\begin{table} \label{tab:exam}
	\caption{Poisson model found through the hs algorithm. bic: 6283.28 Log-Likelihood: -3084.44}
	\begin{center}
		\resizebox{0.45\textwidth}{!}{
        \begin{tabular}{|l|l|c|c|c|l|}
			\hline
			Effect & $\tau$ & Coeff & Std. Err & z-values & Prob |z|>Z \\
			\hline
			const & no & 0.0712 & 1.64 & 0.04 & 0.97 \\
			\hline
			MIGRADE & no & 0.0904 & 0.46 & 0.20 & 0.84 \\
			\hline
			TANGENT & no & -0.4529 & 0.24 & -1.88 & 0.06. \\
			\hline
			FRICTION & nil & -0.3800 & 0.09 & -4.39 & 0.00*** \\
			\hline
			AVESNOW & nil & -0.1213 & 0.19 & -0.65 & 0.52 \\
			\hline
			INCLANES & no & 0.1162 & 0.46 & 0.26 & 0.80 \\
			\hline
			SPEED & sqrt & 0.1287 & 0.76 & 0.17 & 0.87 \\
			\hline
			DOUBLE & no & 0.2838 & 0.12 & 2.35 & 0.02* \\
			\hline
			PEAKHR & nil & 0.3417 & 0.23 & 1.46 & 0.15 \\
			\hline
			ACCESS & no & -0.7397 & 0.95 & -0.78 & 0.44 \\
			\hline
			SLOPE & no & -0.0434 & 0.36 & -0.12 & 0.90 \\
			\hline
			INTECHAG & no & 0.1827 & 0.40 & 0.46 & 0.65 \\
			\hline
			AVEPRE & log & -0.3353 & 1.07 & -0.31 & 0.75 \\
			\hline
			INCLANES (Std. Dev.) uniform &  & 0.2545 & 0.21 & 1.21 & 0.23 \\
			\hline
			SPEED (Std. Dev.) uniform &  & 0.2561 & 0.07 & -3.42 & 0.00*** \\
			\hline
			DOUBLE (Std. Dev.) triangular &  & 0.2003 & 0.12 & 1.71 & 0.09. \\
			\hline
			PEAKHR (Std. Dev.) tn\_normal &  & 2.5883 & 0.06 & -46.68 & 0.00*** \\
			\hline
			ACCESS (Std. Dev.) ln\_normal &  & 0.1714 & 0.20 & 0.87 & 0.38 \\
			\hline
			SLOPE (Std. Dev.) triangular &  & 0.1151 & 0.81 & 0.14 & 0.89 \\
			\hline
			INTECHAG (Std. Dev.) triangular &  & 0.0200 & 0.65 & -0.03 & 0.98 \\
			\hline
			AVEPRE (Std. Dev.) normal &  & 0.0567 & 0.25 & -0.23 & 0.82 \\
			\hline
		\end{tabular}
        }
	\end{center}
\end{table}

\FloatBarrier
\subsection{Initial Solution Search} \label{lab:constaintssection}
The primary purpose of this software is to provide analysts with the capability to traditionally fit models while also exploring alternative solutions. As such, offering an initial solution is recommended. As noted earlier, generating an unguided first solution often results in outcomes with numerous variables that must be rejected. To alleviate the burden on the metaheuristic process, providing an initial solution helps place the search in a favorable starting state, allowing it to be refined and optimized. Consider the initial solution as the "typical case" when analysts fit models and are reasonably satisfied with the initial result.

The following code snippet in \cref{lis:prespec} illustrates how an initial solution can be defined. This includes a set of correlated random solutions generated using the Poisson distribution, with parameters drawn from uniform, normal, and triangular random distributions.

\begin{lstlisting}[language=Python, caption ={Pre-specifying an initial solution.}, label={lis:prespec}] 
 manual_fit_spec = {
'fixed_terms': ['const'],
'rdm_terms':  [],
'rdm_cor_terms': ['X1:uniform', 'X2:normal', 'X3:triangular'],
'grouped_terms': [],
'hetro_in_means': [],
'transformations': ['no', 'no', 'no', 'no'],
'dispersion': 0
}
args['Manual_Fit'] = manual_fit_spec
\end{lstlisting} 
\FloatBarrier
Then, supply the arguments dictionary to the objective function. This will effectively incorporate one of the solutions. This can be observed in \cref{lis:obj}. Note this will estimate a singular model. If instead the \code{obj\_fun} was embed into a search the initial solution specified from \cref{lis:prespec} would be included within the optimization search
\begin{lstlisting}[language=Python, caption = Pushing an initial solution to the objective function to estimate the singular model.] 
 obj_fun = ObjectiveFunction(X, y, **args)

\end{lstlisting} \label{lis:obj}
\FloatBarrier

\subsection{Termination Criterion}
Termination criterion can occur in multiple ways:
\begin{enumerate}
    \item Max time reach
    \item Number of iterations without improvement
    \item Max iterations
\end{enumerate}

All algorithms will terminate immediately upon the first instance of exceeding the time limit given by the \code{\_max\_time} argument in the \code{ObjectiveFunction}, or the iteration limit specified by the \code{\_max\_iter} argument in the retrospective algorithm, or number of iterations without an improving solution\code{\_WIP}. Upon termination, the best solution(s) will be reported, and a Pareto frontier will be plotted if multiple objectives exist.

After termination, the user will receive a list of Pareto optimal solutions, along with the following performance metrics. An en example output can be seen below:
\begin{figure}
    \centering
    \label{fig:enter-label}
\begin{lstlisting}{language = console}
Elapsed time: 0:01:34.087934
Pareto Solutions: [{'aic': 2168.614, 'bic': 2225.0140,...}]
\end{lstlisting}
\caption{Output With Early Termination Based on Time}
\end{figure}
\FloatBarrier
    
\begin{figure}[!h]
    \centering
   \includegraphics[width=.45\textwidth]{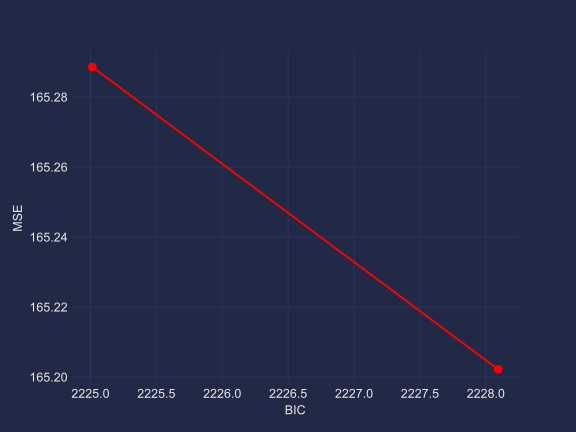}
    \caption{Pareto-efficiency plot returned after the termination criterion.}
    \label{fig:enter-labelpaper4}
\end{figure}

    
\FloatBarrier
The user will also receive a graphical representation of a Pareto frontier. This illustrates all potential solutions that cannot be strictly deemed more efficient than one another with respect to the objective functions, without causing a deterioration in at least one of them. For instance, consider the following output of a multi-objective search for a model as presented in  \cref{fig:enter-labelpaper4}. The model yields two distinct solutions, which, according to the Bayesian Information Criterion (BIC) and the Mean Squared Error (MSE), cannot be considered strictly superior to one another. This offers analysts a list of noncompromising solutions to assess and derive insights from. Depending on the characxtxristics of the problem, numerous additional compromising solutions could have been identified and suggested.
\FloatBarrier
There exist further termination criteria within all algorithms. These criteria revolve around the absence of improvements in the solution arguments. The default parameter \code{arg: - \_max\_iterations\_improvement = 50} can be modified. If no solutions are accepted for consideration in the search process by any algorithm, which includes the potential acceptance of a worse solution, then the search will terminate after 50 iterations. If any solution is found that updates the population, is accepted for consideration, or is pareto-efficient, then the iteration count will be reset.

\section{Hierarchical Capabilities}
\subsection{Available Distributions}
\begin{figure}[!h]
\centering
\resizebox{0.4\textwidth}{!}{
\begin{tikzpicture}
\begin{axis}[
height=10cm,
width=10cm,
axis lines=left,
xlabel=$x$,
ylabel=$f(x)$,
xmin=-4,
xmax=4,
ymin=0,
ymax=1,
xtick={-2, 0, 2},
ytick={0.0, 0.5, 1},
yticklabels={},
xticklabels={$-2$, $0$, $2$},
legend style={at={(0.02,0.98)}, anchor=north west},
legend cell align=left,
clip=true
]

\addplot[blue, domain=-4:4, samples=100] {1/sqrt(2*pi)*exp(-0.5*(x)^2)};
\addlegendentry{Normal}

\addplot[red, domain=-4:4, samples=100] {0.25};
\addlegendentry{Uniform}

\addplot[green!70!black, domain=-4:4, samples=100] {1-abs(x)/2};
\addlegendentry{Triangular}

\addplot[yellow!70!black, domain=0.01:4, samples=100] {1/(x*sqrt(2*pi))*exp(-0.5(ln(x))^2)};
\addlegendentry{Log-Normal}

\addplot[orange!70!black, domain=-4:4, samples=100] {1/(sqrt(2*pi)*0.5)*exp(-0.5(x)^2)};
\addlegendentry{Truncated Normal}


\end{axis}
\end{tikzpicture}
}
\caption{Possible distributional assumptions capable for the random parameters.}
 \label{fig:distributions}
\end{figure}
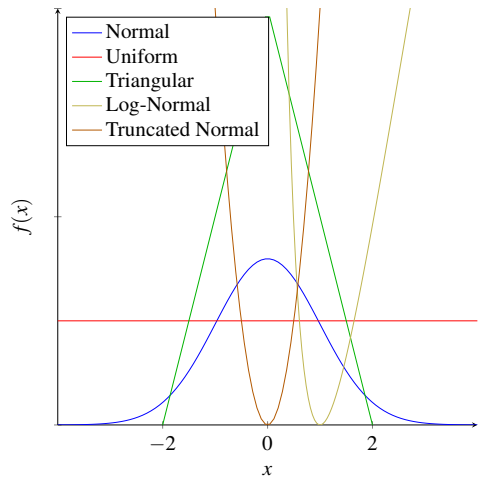

To introduce heterogeneity within the contributing factors, sampling is often performed using a normal distribution, which is commonly the only one tested \citep{Train2009DiscreteSimulation}. To address this limitation, the Metacountregressor software provides a wide range of available distributions. These distributions can be incorporated to capture the variability of contributing factors within standard deviations and account for unobserved heterogeneity. The figure in \cref{fig:distributions} showcases the properties of the distributions available to an analysis in the software for the random parameters.

\FloatBarrier
\subsection{Grouped Random Parameters}
Grouped random parameters are only considered if the \code{ObjectiveFunction} is supplied the argument \code{(-groups = "group_name")}. In this argument, \code{"group_name"} should correspond to the named header of the group elements in the explanatory matrix \code{X}. Please note that grouped random parameters will only be taken into account in cases the user has specifically defined the groups, which should be factor elements. Within the \pkg{pandas} data-frame declare the group. Defining the group will allow the random parameters to be considered together or separately with respect to this grouping element. Groups need to be defined up front. Take data from Maine county for example, in this example we will allow for a decision to group the random parameters by each county.

The following listing shows how to do this.

\begin{lstlisting}[language=Python, caption = Pushing an initial solution to the objective function to estimate the singular model, label= {listi:grp_}] 

 model_terms  = {
            'Y': 'crashes',  # Dependent variable
            'group': 'county',  # Grouping column (if any)
            'panels': 'element_ID',  # Panel column (if any)
            'Offset': None  # Offset column (if any)
        }
args['model_terms] = model_terms
obj_fun = ObjectiveFunction(X, y, **args)
\end{lstlisting} 
\FloatBarrier
Similarly, \cref{listi:grp_} shows the panels and offset. Defining these labels treats the named columns in \code{X} as the grouping column, panel colum, or offset column.

\subsection{Panel Data}
Panel data will only be taken into account if the panel arguments are specified as \code{(-panels = "panel_id")}. Here, \code{"panel_id"} should correspond to the named header of the group elements in the explanatory matrix \code{X}. When the panels are designated, the assumed format for estimating models is cross-sectional \citep{Greene2007FunctionalData}. Otherwise, if panels are not specified, models are estimated with each observation treated independently. The format of the panel data is detailed below, with the column `panel\_id` designating the unique identifier for each panel. Note that the panels do not need to be balanced; the number of time periods can vary across panels. Below is a sample table illustrating this structure:

\begin{table}[h!]
    \centering
    \begin{tabular}{|c|c|c|c|}
        \hline
        \textbf{panel\_id} & \textbf{time} & \textbf{...} & \textbf{...} \\ \hline
        1                 & 1             & ...          & ...          \\ 
        1                 & 2             & ...          & ...          \\ 
        2                 & 1             & ...          & ...          \\ 
        2                 & 3             & ...          & ...          \\ 
        3                 & 1             & ...          & ...          \\ 
        3                 & 2             & ...          & ...          \\ 
        3                 & 3             & ...          & ...          \\ \hline
    \end{tabular}
    \caption{Structure of Panel Data Highlighting `panel\_id`}
    \label{tab:panel_data}
\end{table}

\subsection{Heterogeneity in the Means} \label{sec:het}
Another feature of the \pkg{MetaCountRegressor} package is its ability to capture heterogeneity in the means of the random parameters. This is an important feature because it allows to account for variability in the means, which is distinct from standard random parameters. The key difference is that heterogeneity in parameter for a road attribute is often a function of the other spatial characteristics, rather than being influenced solely by human-related or environmental factors. To address this, the mean of the parameter distribution for individual road segments can be expressed as a function of other observed road characteristics using a random parameters approach with heterogeneity in means.

To enable heterogeneity in the means, as a feature within the search, a user can use the complexity\_level=6 \footnote{complexity\_level=6 $\rightarrow$ complexity\_level= $[0,1,3,4,5]$} parameter when initializing the \pkg{MetaCountRegressor} package. This allows the model to estimate random parameters to capture the unobserved variation in the model means.


Furthermore, the user can also setup an initial solution that capture heterogeneity in the means. Consider the following example on how to model for the heterogeneity in means between Z1 and X2 and X3. See \cref{listing:2}

\begin{lstlisting}[language=Python, caption = {Heterogeneity in the means specification, example syntax.}] 
manual_fit_spec = {
'fixed_terms': ['const'],
'rdm_terms':  [ 'X1:uniform'],
'rdm_cor_terms': [],
'grouped_terms': [],
'hetro_in_means': ['Z1:normal', 'X2:normal', 'X3:normal'],
'transformations': ['no', 'no', 'no', 'no', 'no', 'no'],
'dispersion': 0
}
args['Manuel_Fit'] = manual_fit_spec
obj_fun = ObjectiveFunction(X, y, **arguments)
\end{lstlisting} \label{listing:2}

The output of this model can be seen in \cref{table:h}.
\begin{figure}
\centering
\begin{lstlisting}[
    basicstyle=\ttfamily\tiny,        % Monospaced font with small size
    backgroundcolor=\color{black!5},   % Light gray background
    frame=single,                      % Single-line frame
    showstringspaces=false,            % Don't show spaces in strings
    breaklines=true,                   % Line breaking
    columns=fullflexible               % Better alignment for monospaced font
]
--------------------------------------------------------------------------------
Log-Likelihood:  -2359.880062678968
--------------------------------------------------------------------------------
bic: 4768.05
--------------------------------------------------------------------------------
+------------------------+--------+---------+----------+----------+------------+
|         Effect         | $\tau$ |  Coeff  | Std. Err | z-values | Prob |z|>Z |
+========================+========+=========+==========+==========+============+
| const                  | no     | 0.3300  |   0.15   |   2.20   | 0.03       |
+------------------------+--------+---------+----------+----------+------------+
| X1                     | no     | 0.5321  |   0.05   |   10.64  | 0.00       |
+------------------------+--------+---------+----------+----------+------------+
| X1 (Std. Dev.) uniform | no     | 0.0558  |   0.05   |  -1.12   | 0.26       |
+------------------------+--------+---------+----------+----------+------------+
| main: X2: hetro group  | no     | -0.3192 |   0.17   |  -1.88   | 0.06       |
| 0                      |        |         |          |          |            |
+------------------------+--------+---------+----------+----------+------------+
| X3: hetro group 0      | no     | 1.4169  |   0.03   |   47.23  | 0.00***    |
+------------------------+--------+---------+----------+----------+------------+
| Z1: hetro group 0      |        | 0.4249  |   0.09   |   4.72   | 0.00***    |
+------------------------+--------+---------+----------+----------+------------+
| main: X2: hetro group  |        | 0.1944  |   0.00   |   50.0   | 0.00***    |
| 0:normal:sd  hetro     |        |         |          |          |            |
| group 0                |        |         |          |          |            |
+------------------------+--------+---------+----------+----------+------------+
\end{lstlisting}
\caption{Output from heterogeneity in the means}
\label{table:h}
\end{figure}
\FloatBarrier
\section{Hyper-Parameter Settings}
\subsection{Algorithm Specific Hyperparameters}
The following code in \cref{listing:hyper} allow the user to select the hyperparamaters for each algorithm of the search. 
\begin{lstlisting}[language=Python, caption = {Defining different arguments for the solution algorithm.}, label = {listing:hyper}] 
'''Arguments for the solution algorithm'''
argument_hs = {
    '_hms': 20, #harmony memory size,
    '_mpai': 1, #adjustement inded
    '_hmcr': .5
}
obj_fun = ObjectiveFunction(X, y, **arguments)
results = harmony_search(obj_fun, None, argument_hs)
print(results)


argument_de = {'_AI': 2,
            '_crossover_perc': .2,
            '_max_iter': 1000,
            '_pop_size': 25
}
de_results = differential_evolution(obj_fun, None, **argument_de)
print(de_results)


args_sa = {'alpha': .99,
        'STEPS_PER_TEMP': 10,
        'INTL_ACPT': 0.5,
        '_crossover_perc': .3,
        'MAX_ITERATIONS': 1000,
        '_num_intl_slns': 25,
}
sa_results = simulated_annealing(obj_fun, None, **args_sa)
print(sa_results)
\end{lstlisting} 
Because hyperparameters significantly influence the effectiveness of the search process, testing was conducted to identify default configurations that would allow the algorithms to work effectively "out of the box" for analysts on first use. To achieve this, we performed experiments to determine appropriate default values, as detailed in Section~\ref{submulti}. These experiments involved testing the algorithms Simulated Annealing (SA), Harmony Search (HS), and Differential Evolution (DE), each with different hyperparameter settings. The experiments explored all possible combinations of the following hyperparameter grids:

\begin{itemize}
    \item SA:
The hyperparameter grid for SA is as follows:
\begin{itemize}
    \item \texttt{alpha}: Cooling rate values \([0.9, 0.925, 0.95, 0.975, 0.99]\)
    \item \texttt{STEPS\_PER\_TEMP}: Steps per temperature \([5, 10, 20]\)
    \item \texttt{INTL\_ACPT}: Initial acceptance rate: \([0.5]\)
    \item \texttt{\_crossover\_perc}: Crossover percentage \([0.2, 0.3, 0.45, 0.6, 0.7]\)
    \item \texttt{\_num\_intl\_slns}: Number of initial solutions: 25
\end{itemize}

\item DE:
The hyperparameter grid for DE is as follows:
\begin{itemize}
    \item \texttt{\_AI}: Adjustment index \([1,2,3]\)
    \item \texttt{\_crossover\_perc}: Crossover percentage \([0.2, 0.3, 0.4, 0.5, 0.6, 0.7]\)
    \item \texttt{\_pop\_size}: Population sizes \([10, 20, 30, 40, 50]\)
\end{itemize}

\item HS
The hyperparameter grid for HS is as follows:
\begin{itemize}
    \item \texttt{\_hms}: Harmony memory sizes \([10, 20, 30, 50]\)
    \item \texttt{\_hmcr}: Harmony memory consideration rate \([0.3, 0.45, 0.55, 0.7]\)
    \item \texttt{\_mpai}: Minimum pitch adjustment index \([1, 2,3]\)
\end{itemize}
\end{itemize}
By systematically testing these parameter combinations, it was aimed to identify how sensistive these configurations are across a range of scenarios. The results of these experiments, are discussed further in Section~\ref{submulti}, where 5 seeded runs were performed using the hetrogeneity in the means dataset from Section ~\ref{sec:het}.



\subsection{Multiple Objective Settings} \label{submulti}
The results of the experiments can be seen in \cref{algpap4:hs,algpap4:sa,algpap4:de}, which represent 5 seeded runs for each hyperparamater configuration.
In this dataset where multiple objectives are considered, \gls{HS} appears to be the most effective approach. The smaller weighted sums of both objectives across the problem instance with varying hyperparameter settings indicate that \gls{HS} is highly robust and performs consistently well. This suggests that \gls{HS} is relatively insensitive to hyperparameter changes, making it a reliable choice for multi-objective optimization tasks.

While \gls{HS} displayed slight sensitivity to smaller population sizes, its overall performance remained stable, indicating that fine-tuning this parameter has a limited impact on its efficiency. Such robustness to hyperparameter variations is a significant advantage, especially in scenarios where computational resources or time for extensive hyperparameter tuning are limited.

All algorithms evaluated in this study performed relatively well the tested problem instance. However, the adaptability and ability of \gls{HS} to consistently achieve lower weighted sums of objectives, even under varying hyperparameter configurations, make it an effective approach for solving complex multi-objective optimization problems. 

\begin{figure}[htbp]
\centering
\includegraphics[width = .45\textwidth]{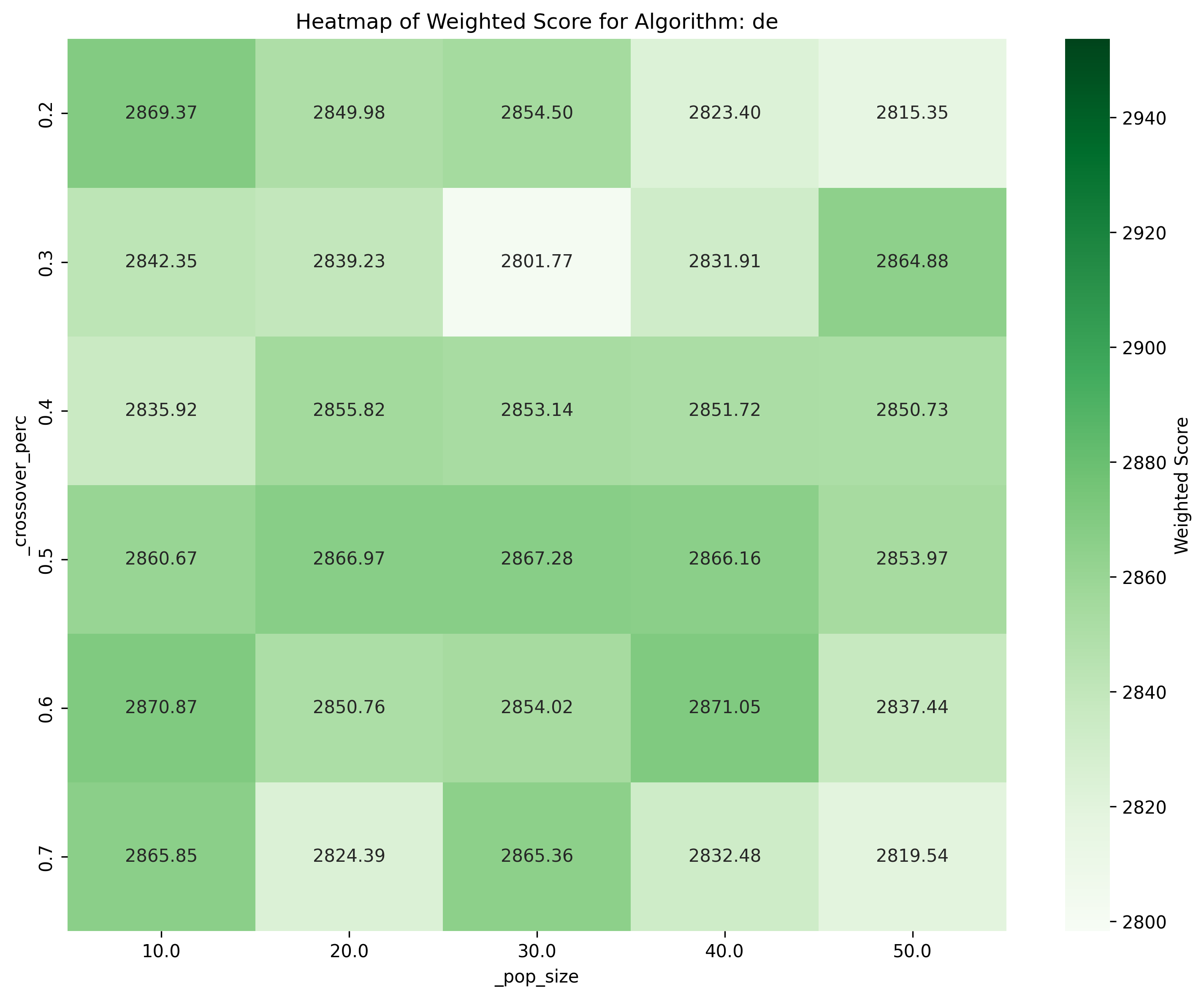}
\caption{Covering Arrays minimizing the weighted sums for differential evolution.}
\label{figpap4:de}
\end{figure}
\FloatBarrier
\begin{figure}[htbp]
\centering
\includegraphics[width=.45\textwidth]{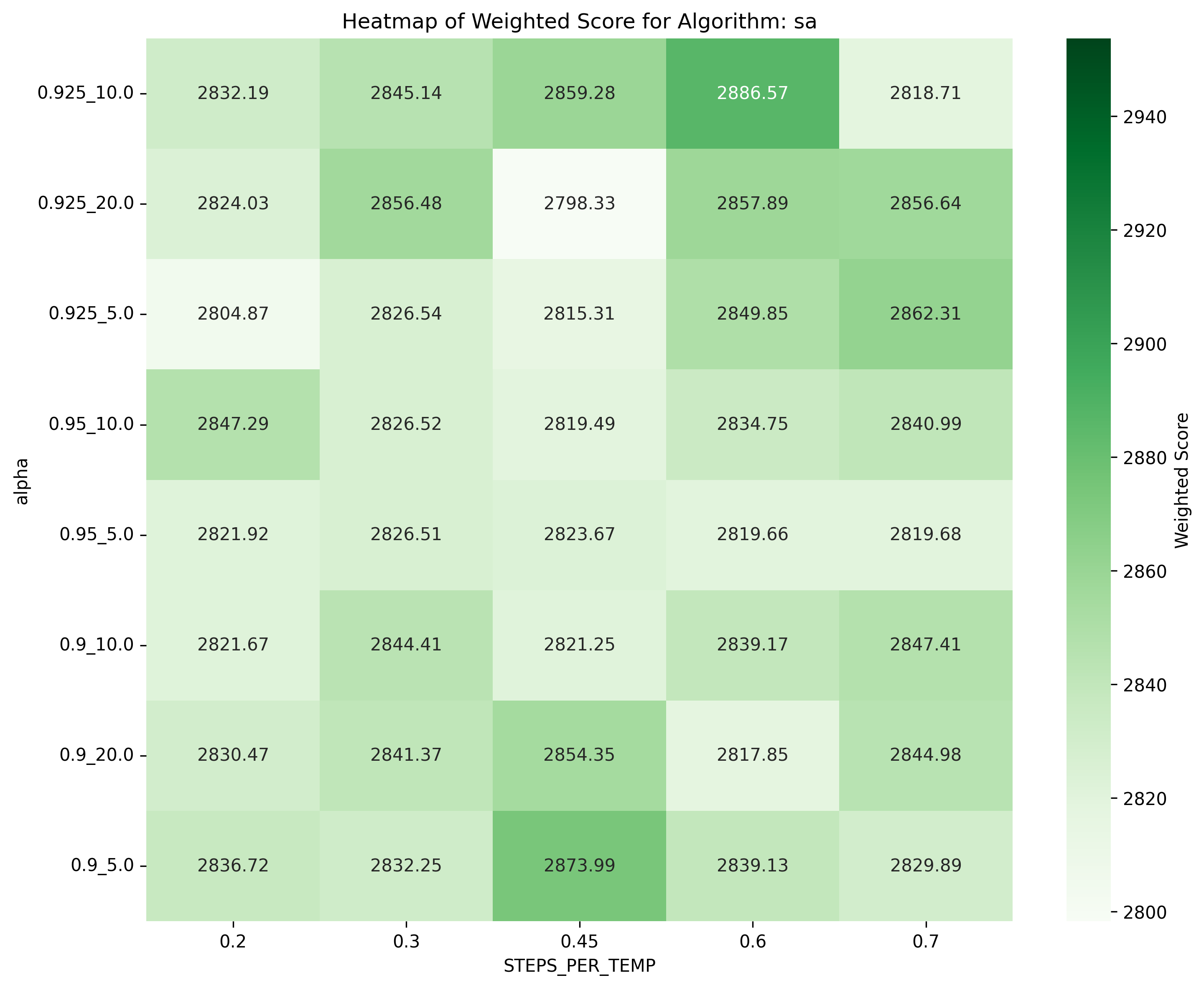}
\caption{Covering Arrays minimizing the weighted sums for simulated annealing.}
\label{figpap4:sa}
\end{figure}
\noindent
\begin{figure}[htbp]
\centering
\includegraphics[width = .45\textwidth]{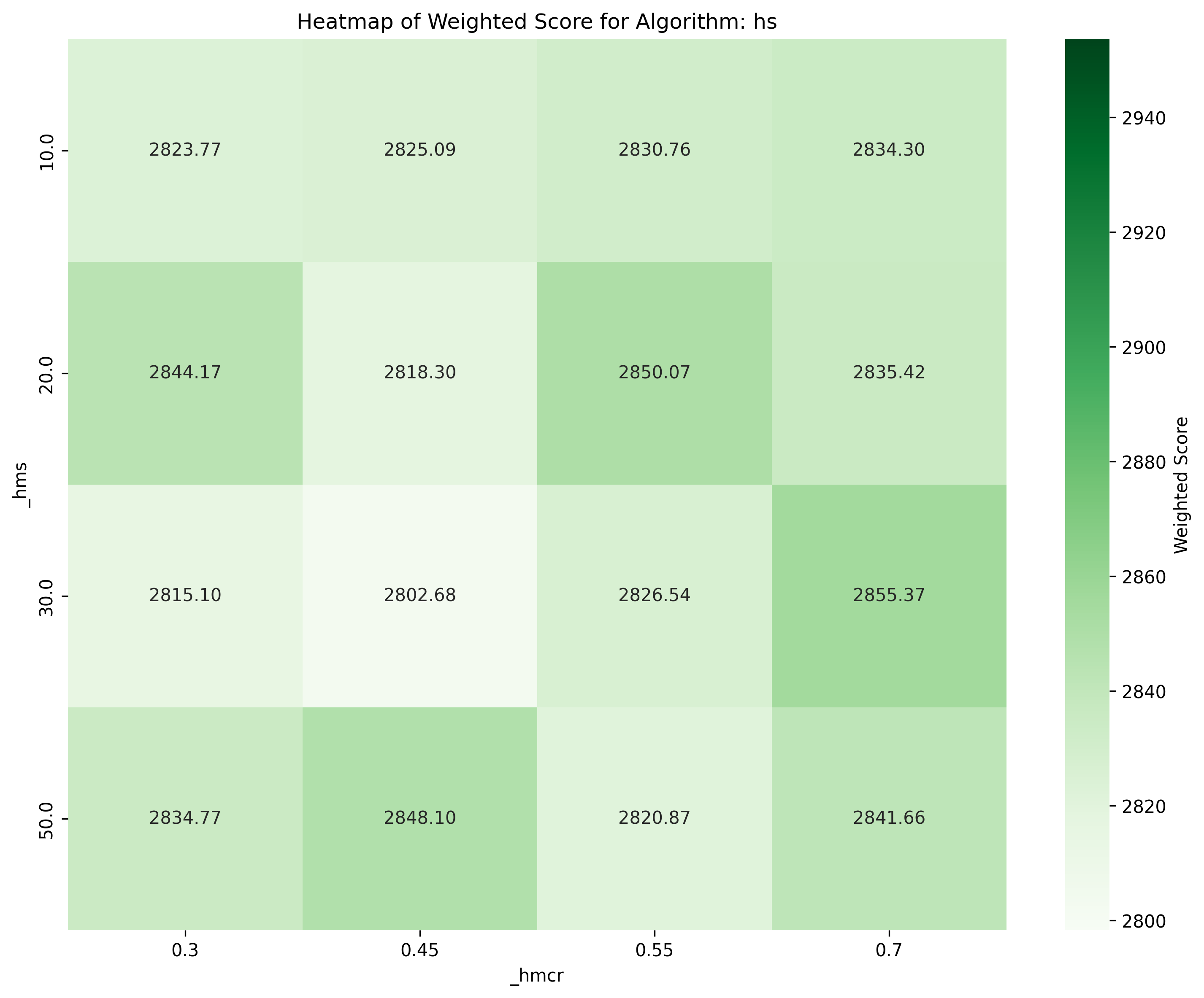}
\caption{Covering Arrays minimizing the weighted sums for harmony search.}
\label{figpap4:hs}
\end{figure}
\noindent
\begin{figure}[htbp]
\centering
\includegraphics[width = .45\textwidth]{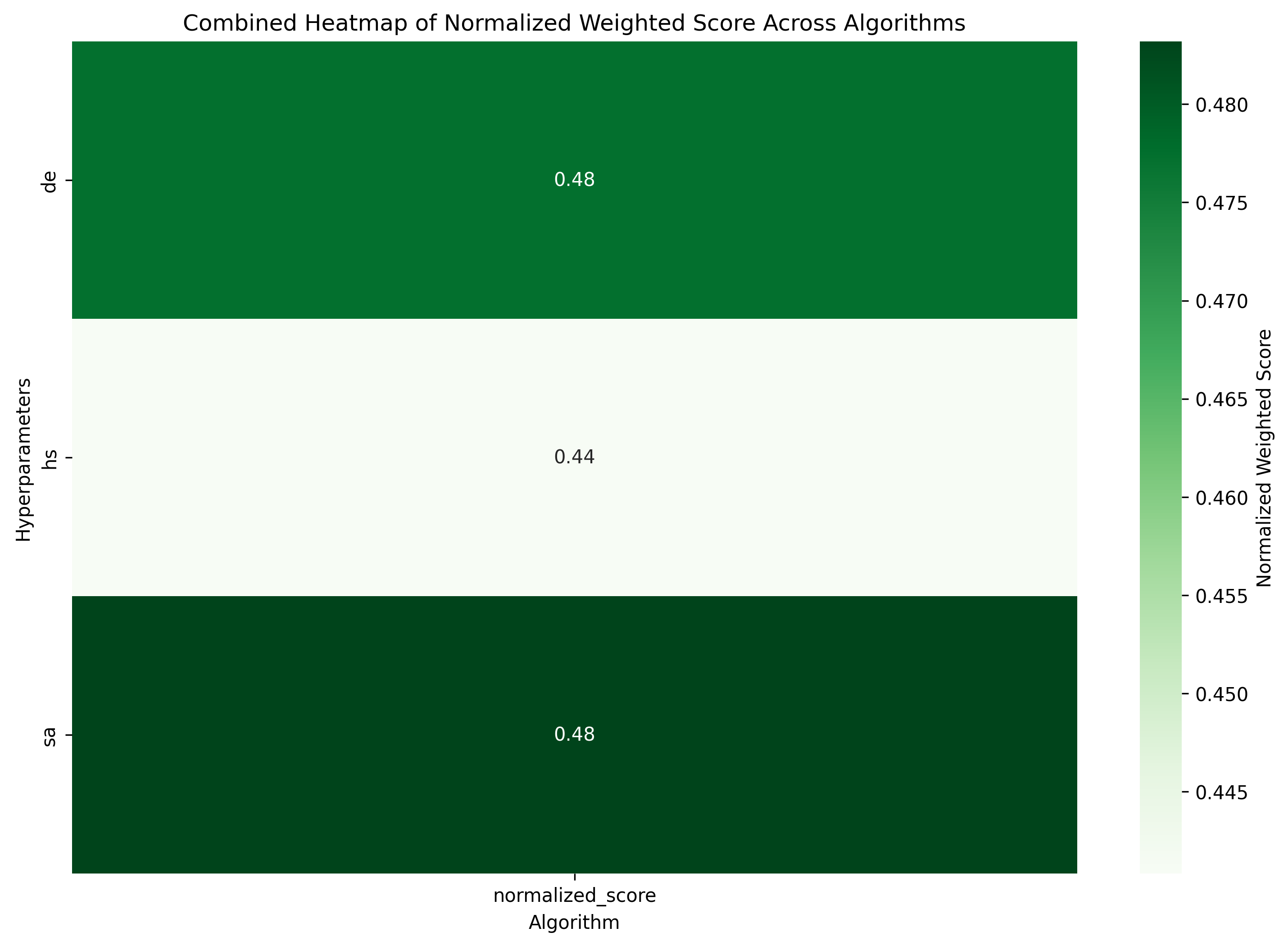}
\caption{Covering Arrays for Compared}
\label{figpap4:all}
\end{figure}
\FloatBarrier

Note that the results shown in \cref{figpap4:all} relate to this specific experiment conducted on a single dataset. Different results may be observed when using datasets with significantly different characteristics, observations, or objectives for the search itself.
\section{Analyst Use Case}
The framework is designed to incorporate as much information as possible, allowing models to be searched with minimal analyst experience. In \cref{lab:constaintssection}, an initial solution was embedded into the framework. A similar test is now being considered, but with a focus on restricting certain hypotheses to be tested.

The variables described in \cref{secpap4:variables} are considered. The variable X1 is assigned six levels of freedom: off (level 0), fixed effects (level 1), random parameters (level 2), correlated random parameters (level 3), grouped random parameters (level 4), and heterogeneous means (level 5). In this context, each model search is restricted to testing one of these attributes for "X1."

Similarly, "X2" is defined at levels 1, 2, or 5, meaning it is constrained such that every potential model includes this variable, as level 0 is not defined. However, a choice is still provided to test whether random parameters or heterogeneity in the means are required.

Lastly, "X3" is restricted to operate at levels 1, 3, or 5, meaning it is only tested as a fixed effect, a correlated random parameter, or with heterogeneous means.

The role of distributions is governed by the same logic as level specifications. Specifically, for levels 0 and 1, the distributions have no impact on the search, as they become relevant only when random parameters are considered. Moreover, by defining a specific set of distributions, only those within the set are considered. For example, for "X3," only normal or lognormal distributions are tested.
\begin{lstlisting}[language=Python, caption = {Enforcing variable constriants}, label = {secpap4:variables}] 
#dictionary of variables in the dataset with freedom to explore options.

variable_decisions = {
    'X1': {'levels': [0,1], 'transformations': ['no'], 'distributions': []},
    'X2': {'levels': [1,2,5], 'transformations': ['no'], 'distributions': ['n', 't']},
    'X3':{'levels': [0,2,6], 'transformations': ['no'], 'distributions': ['n', 'ln']},
    'Z1': {'levels': [0,5], 'transformations': ['no'], 'distributions': ['n']},
    'Z2': {'levels': [0,2,5], 'transformations': ['no'], 'distributions': ['ln']}
    }

a_des, X = helperprocess.set_up_analyst_constraints(X, model_terms, variable_decisions)
arguments['decisions'] = a_des
obj_fun = ObjectiveFunction(X, y, **arguments)
results = harmony_search(obj_fun)
helperprocess.results_printer(results, arguments['algorithm'], int(arguments['is_multi']))
\end{lstlisting} 
To evaluate this setup, we compare an iteration plot of the proposed configuration against an alternative where all decisions are given complete freedom to be explored. This comparison shows that convergence is achieved much faster with the proposed setup. The proposed setup was designed with knowledge of the data generation process. For example, we already know that this model should exhibit heterogeneity in the means with respect to Z1. However, restricting this to test only for heterogeneity in the means or turning it off has significantly reduced the number of decisions. Furthermore, the distributions tested in this setup are limited to normal distributions. This ensures that the true specification remains within the tests, while many potential false alternatives are excluded from consideration.

\begin{figure}[ht]
    \centering
    \begin{subfigure}[b]{0.9\linewidth}
        \centering
        \includegraphics[width=\linewidth]{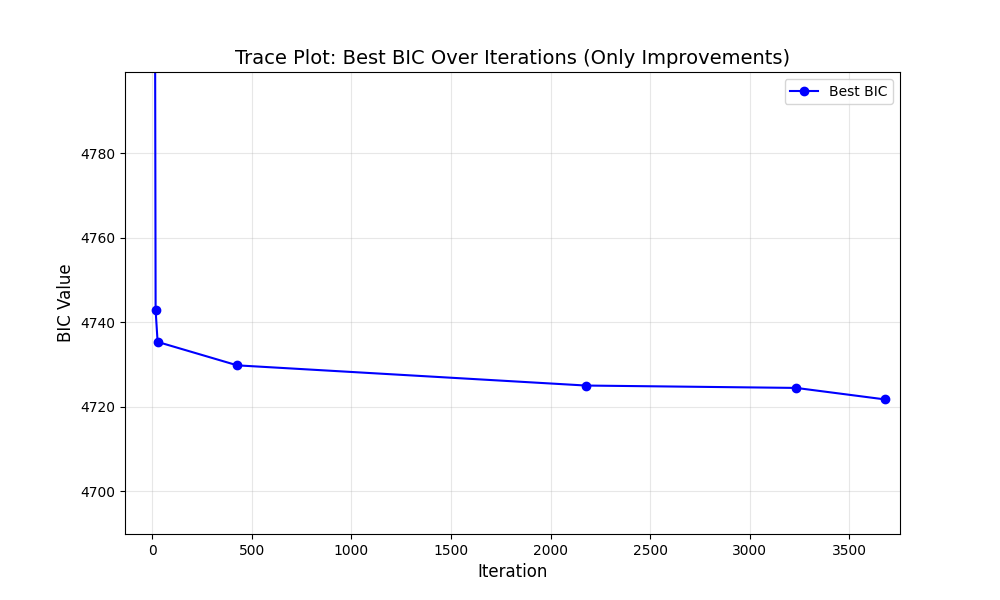}
        \caption{Unconstrained}
        \label{subfig:unconstrained}
    \end{subfigure}%
    \hfill
    \begin{subfigure}[b]{0.9\linewidth}
        \centering
        \includegraphics[width=\linewidth]{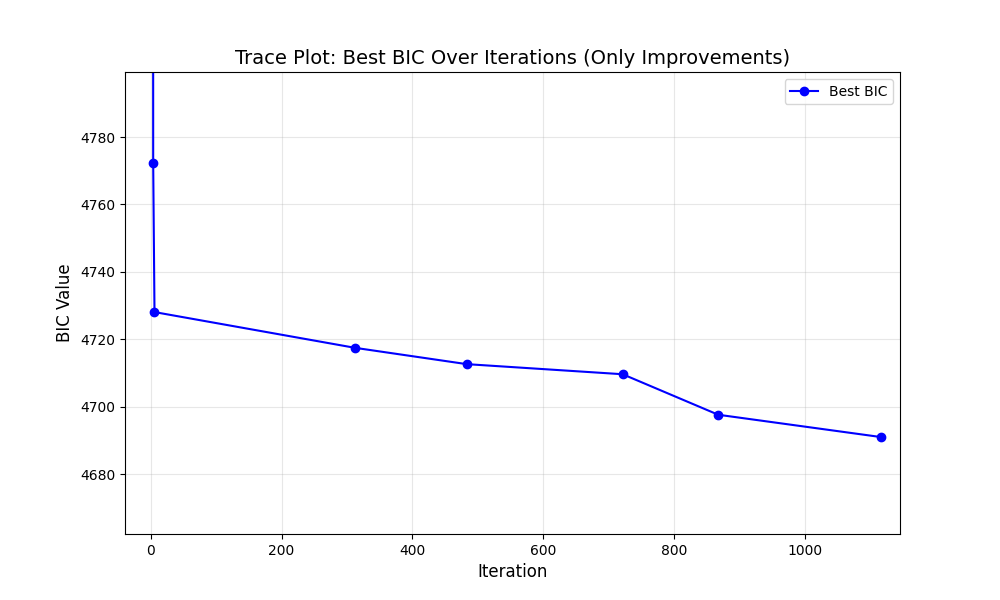}
        \caption{Constrained}
        \label{subfig:constrained}
    \end{subfigure}
    \caption{Comparison of unconstrained and constrained cases.}
    \label{fig:comparison}
\end{figure}
\FloatBarrier
Observing \Cref{fig:comparison}, it can be seen that introducing intuitive constraints to the problem, such as limiting the number of tests for specific types of distributions, helps to reduce the search space and identify solutions more efficiently in terms of the required iterations. These constraints should be applied in conjunction with the analyst's knowledge and expertise. While this approach can improve optimization in terms of computation time, there is also a risk that an analyst might inadvertently exclude a potential test that could have led to a better model.

\section{Summary and discussion} \label{secpap4:summary}
This paper aims to provide valuable insights into the properties and functionalities of the \pkg{MetaCountRegressor} software package for the analysis of count data. Its primary objective is to facilitate the efficient development of models that meet specific data criteria while assisting analysts in identifying relevant features through an optimization solution algorithm. In order to achieve this, analysts must input parameters and constraints to customize the search process, leveraging their knowledge of the underlying modeling framework to maximize software capabilities.

Although the default arguments of the software offer an efficient starting point for data exploration, it is crucial to gather as much information as possible from the analysts to effectively guide the search. Restricting the number of distributions or limiting the hierarchical structure of the model's capabilities, for example, can be useful in aligning with analysts' preferences while still benefiting from the efficiency of the solution algorithms. This approach ensures that the analyst's role is not replaced by the framework but rather augmented, providing additional support for the complex testing required. Using the framework as a decision-assisted tool, analysts can efficiently and effectively test large numbers of hypotheses that might otherwise have been overlooked. This collaborative approach combines the strengths of the framework with the expertise and knowledge of the analyst, resulting in a more comprehensive and insightful analysis.

To complement this paper, a tutorial in the form of a supplementary Jupyter notebook is provided. This notebook serves as an invaluable resource, offering clearer instructions and intuitive examples on how to effectively use the package as an open resource.

By creating this tutorial and making the software accessible to all, we aim to promote research endeavours and encourage the adoption of new capabilities and improvements. This inclusive approach will contribute to a more efficient and reliable modeling process, benefiting a wide range of users.


\appendix
\section*{Computational details}

The results in this paper were obtained using
\proglang{Python}~3.10.1 with the
\pkg{MetaCountRegerssor}~1.0.04 package. \proglang{Python} itself
and all packages used are available from the
\proglang{Python} PyPi package directory  (CRAN) at
\url{https://CRAN.Python-project.org/MetaCountRegressor}.

\section*{Data Availablility}
Data openly available in a public repository that does not issue DOIs. This can be found at \url{https://anonymous.4open.science/r/data-8257}.

\onecolumn
\section{Algorithm Pseudocode}
\Cref{tab:meta4} details the notation used for the algorithms, as well as the experimental parameters and notation concerning the solution algorithm.

\begin{longtable}[H]{K{0.15\linewidth}K{0.73\linewidth}}
\caption{Notation: Representation of the solution and the properties of the metaheuristics.}
\label{tab:meta4} \\
\toprule
\multicolumn{2}{c}{Experimental Default Parameters} \\ \midrule
$MaxTime$ & Max allowable time for each algorithm.\\ 
\midrule
\multicolumn{2}{c}{Solution Representation} \\ \midrule
$\mathcal{P}$ & Pareto-efficient set of solutions \\
$v_p$ & Solution vector to the mathematical program, where $s$ depicts a solution for population at indexed p. \\
$v_{p,j}$ & Specification of decision element $j \; \forall j \in \mho$.\\
$\ell_{p, j}$ & Index value of the solution $p$ for decision $j$.\\
\midrule
\rowcolor{white}\multicolumn{2}{c}{Harmony Search} \\ \midrule
$PAI$ & Pitch adjustment index. \\
$HM$ & Set of distinct solutions ($v_p$) in the harmony memory.\\
$HMS$ & Harmony memory size (controls the set size of $HM$).\\
$HMCR$ & The probability of retaining the previous solution\\ \midrule
\multicolumn{2}{c}{Differential Evolution} \\ \midrule
$CR$ & The probability of a crossover occurring on $v_p$\\ 
$PS$ & The population size. \\ \midrule
\rowcolor{white}\multicolumn{2}{c}{Simulated Annealing} \\ \midrule
$T_{step}$ & Number of iterations before the temperature is reduced.\\
$T_{\alpha}$ & Temperature reduction factor.\\
\bottomrule
\end{longtable}	
\section{Algorithmic logic} \label{alg:exppap4}
\Cref{algpap4:de,algpap4:sa,algpap4:hs} presents the pseudocode for the metaheuristics that guide the search for a regression model.
\\

\begin{algorithm} 
\renewcommand\footnoterule{}
\begin{algorithmic}[1] 
\Function{MOGBHS}{$HMS$, $HMCR$, $PAI$, $MaxTime$}
\State HM $\leftarrow$ \textproc{Generate\_Random\_Specifications}(HMS) \footnotemark \Comment{Generate initial population.}
\State HM $\leftarrow$ \textproc{SPEA2\_Sort}(HM) \footnotemark \Comment{Sort using SPEA2.}
\State $\mathcal{P}$ $\leftarrow$ \textproc{NonDominatedSort}(HM) \footnotemark \Comment{Track Pareto Set.}
\While{\textproc{current\_runtime}() $\leq$ $MaxTime$}
\For {$j = 1, 2, \dots, HMS$}
\If{$U(0,1) \leq HMCR$}
\State r $\leftarrow$ \textproc{rand\_int}(1, HMS) \footnotemark \Comment{Select random specification.}
\State H[j] $\leftarrow$ HM[r, j]
\Else
\State H[j] $\leftarrow$ \textproc{rand\_decision}(j) \footnotemark[4] \Comment{Generate random specification.}
\EndIf
\If{$U(0,1) \geq PAR$}
\State $H[j] \leftarrow$ \textproc{Conduct\_pitch\_adjustment}() \Comment{Pitch adjustment.}
\EndIf
\EndFor
\State $f_1(\mho), f_2(\mho)$ of H \Comment{Evaluate objectives.}
\If{$f_1(\mho), f_2(\mho) \leq f_1(\mho), f_2(\mho)$ of $(HM[HMS])$}
\State $HM[HMS] \leftarrow H$ \Comment{Replace the worst solution.}
\State $HM \leftarrow$ \textproc{SPEA2\_Sort}($HM$)
\State $\mathcal{P}$ $\leftarrow$ \textproc{NonDominatedSort}($\mathcal{P} \cup H$)
\EndIf
\EndWhile
\EndFunction
\end{algorithmic}  
\caption{Multi-Objective Global Best Harmony Search} \label{algpap4:hs}
\footnotetext[1]{Generates a population of random specifications to the size of the input.}
\footnotetext[2]{Sort by the SPEA2 algorithm.}
\footnotetext[3]{Breaks down to the Pareto Efficient Set, such that each solution is non-dominating to each other.}
\footnotetext[4]{Generates a random continuous value uniformly distributed between 0 and 1.}
\footnotetext[5]{For element $j$ of the chromosome, generate any possible decision randomly available.}
\end{algorithm}

\begin{algorithm}
\renewcommand\footnoterule{}
\begin{algorithmic}[1]
\Function{MOSA}{$T_{0}, T_{end}, MaxTime$, $\chi$, $\chi_n$}
\State $P_0$ = \{$s_1, \ldots, s_{\chi_n} $\} $\gets $ \textproc{Generate\_Random\_Specifications}($\chi_n$)
\State $T \gets T(\chi, P_0)$ 
\State $\mathcal{P} \gets $ \textproc{NonDominatedSort}($P_0$)
\State $s \gets$ \textproc{SPEA2\_Sort}$(\mathcal{P})[1] $\Comment{Gets the best solution from the Pareto}
\State $E_1, E_2 \gets$ $f_1({\mho})$ and $f_2({\mho})$ of $s$ \Comment{Calculate the two objectives}
\While{\textproc{current\_runtime}() $<$ MaxTime}
\For{q = 1,2, $\ldots, T_{\text{step}}$}
\State $s' \gets$ randomly chosen neighbouring state of $s$
\State $E_1', E_2' \gets$ $f_1({\mho})$ and $f_2({\mho})$ of $s'$
 \If{$E_1' < E_1$ or $E_2' < E_2$}
\State $s \gets s'$
\State $\mathcal{P} \gets$ \textproc{NonDominatedSort}($\mathcal{P}\cup s$)
\State $E_1, E_2 \gets E_1', E_2'$
\Else
\State $q \gets e^{-(E_1' - E_1 + E_2' - E_2)/T}$
\If{$\mathcal{U}(0,1) < q$}
\State $s \gets s'$
\State $E_1, E_2 \gets E_1', E_2'$
\EndIf
\EndIf
\EndFor
\State $T \gets$ $T\times T_\alpha$
\EndWhile
\EndFunction
\end{algorithmic}
\caption{Multi-Objective Simulated Annealing Algorithm} \label{algpap4:sa}
\end{algorithm}

 \begin{algorithm}
 \renewcommand\footnoterule{}
\begin{algorithmic}[1]
\Function{MODE}{$CR$, $PS$, $MaxTime$}
\State HM  $\leftarrow$\textproc{Generate\_Random\_Specifications}($PS$)
\State HM $\leftarrow$ \textproc{SPEA2\_Sort}(HM)
\State $\mathcal{P} \gets$ \textproc{NonDominatedSort}(HM)
\While{\textproc{current\_runtime}() $\leq$ $MaxTime$}
\For{$p = 1,2,\ldots,HMS$}
\State $x_p \leftarrow HM[p]$
\State $r_1 \leftarrow$ \textproc{rand\_int}$(1, HMS) :r_1 \neq p$
\State $r_2 \leftarrow$ \textproc{rand\_int}$(1, HMS): r_2 \neq \{p, r_1\}$ 
\State $r_3 \leftarrow$ \textproc{rand\_int}$(1, HMS): r_3 \neq \{p, r_1, r_2\}$
\For{each dimension $j\in  v_p$}
\State $v_{p,j} \leftarrow x_{p,j}{[(\ell_{p, j} +PAI(\ell_{p',j}-\ell_{p'',j}) )\mod(|v_{p,j}|)]} $\Comment{Offspring created.}
\If {$U(0,1) < CR$} 
\State $u_{pj} \leftarrow v_{pj}$ \Comment{Crossover Occurs on element j}
\Else 
\State $u_{pj} \leftarrow x_{pj}$ \Comment{Element j remains the same for solution $x_p$}
\EndIf
\EndFor
\EndFor
\For{$p = 1,2,\ldots,HMS$}
\If{$f_1({\mho}), f_2({\mho})$ of $u_p \leq f_1({\mho}), f_2({\mho})$ of $(HM[p])$ } \Comment{Compare solution to worst in HM} 
\State $HM[p] \leftarrow \mu_p$
\State $HM \leftarrow $\textproc{SPEA2\_Sort}($HM$) \Comment{update population}
\State $\mathcal{P}$ $\leftarrow$ \textproc{NonDominatedSort}($\mathcal{P} \cup u_p$)
\EndIf
\EndFor
\EndWhile
\EndFunction
\end{algorithmic} 
 \caption{Multi-Objective Differential Evolution} \label{algpap4:de} 
\end{algorithm}

\FloatBarrier








\section*{Conflict of interest}

The authors declare no potential conflict of interests.


\begin{thebibliography}{43}
\expandafter\ifx\csname natexlab\endcsname\relax\def\natexlab#1{#1}\fi
\expandafter\ifx\csname url\endcsname\relax
  \def\url#1{\texttt{#1}}\fi
\expandafter\ifx\csname urlprefix\endcsname\relax\def\urlprefix{URL }\fi

\bibitem[{Abdel-Basset et~al.(2018)Abdel-Basset, Abdel-Fatah, and
  Sangaiah}]{Abdel-Basset2018MetaheuristicReview}
Abdel-Basset, M., Abdel-Fatah, L. \& Sangaiah, A.K. (2018) {Metaheuristic
  Algorithms: A Comprehensive Review}. {\it Computational Intelligence for
  Multimedia Big Data on the Cloud with Engineering Applications},,
  185--231.doi:10.1016/B978-0-12-813314-9.00010-4.

\bibitem[{Ahern et~al.(2023)Ahern, Corry, and
  Paz}]{Ahern2023MetacountregressorPyPI}
Ahern, Z., Corry, P. \& Paz, A. (2023) {\it {metacountregressor
  {\textperiodcentered} PyPI}}.
\newline\urlprefix\url{https://pypi.org/project/metacountregressor/0.1.28/}

\bibitem[{Ahern et~al.(2024{\natexlab{a}})Ahern, Corry, and
  Rabbani}]{Ahern2024ExtensiveModels}
Ahern, Z., Corry, P. \& Rabbani, W. (2024) {Extensive hypothesis testing for
  estimation of crash frequency models}. {\it Heliyon}, 10.
  doi:10.1016/j.heliyon.2024.e26634.
\newline\urlprefix\url{http://creativecommons.org/licenses/by/4.0/}

\bibitem[{Ahern et~al.(2024{\natexlab{b}})Ahern, Corry, Rabbani, and
  Paz}]{Ahern2024Multi-objectiveModels}
Ahern, Z., Corry, P., Rabbani, W. \& Paz, A. (2024) {Multi-objective extensive
  hypothesis testing for the estimation of advanced crash frequency models}.
  {\it Accident Analysis {\&} Prevention}, 206, 107690.
  doi:10.1016/J.AAP.2024.107690.

\bibitem[{Akbar et~al.(2024)Akbar, Khan, Shameem, and
  Nadeem}]{Akbar2024GeneticProjects}
Akbar, M.A., Khan, A.A., Shameem, M. \& Nadeem, M. (2024) {Genetic model-based
  success probability prediction of quantum software development projects}.
  {\it Information and Software Technology}, 165, 107352.
  doi:10.1016/J.INFSOF.2023.107352.
\newline\urlprefix\url{https://www.sciencedirect.com/science/article/pii/S0950584923002070}

\bibitem[{Ali et~al.(2022)Ali, Haque, Zheng, and Afghari}]{Ali2022AEnvironment}
Ali, Y., Haque, M.M., Zheng, Z. \& Afghari, A.P. (2022) {A Bayesian correlated
  grouped random parameters duration model with heterogeneity in the means for
  understanding braking behaviour in a connected environment}. {\it Analytic
  Methods in Accident Research}, 35, 100221. doi:10.1016/J.AMAR.2022.100221.

\bibitem[{Beeramoole et~al.(2023)Beeramoole, Arteaga, Pinz, Haque, and
  Paz}]{Beeramoole2023ExtensiveModels}
Beeramoole, P.B., Arteaga, C., Pinz, A., Haque, M.M. \& Paz, A. (2023)
  {Extensive hypothesis testing for estimation of mixed-Logit models}. {\it
  Journal of Choice Modelling}, 47, 100409. doi:10.1016/J.JOCM.2023.100409.

\bibitem[{Behara et~al.(2021)Behara, Paz, Arndt, and
  Baker}]{Behara2021AQueensland}
Behara, K.N., Paz, A., Arndt, O. \& Baker, D. (2021) {A random parameters with
  heterogeneity in means and Lindley approach to analyze crash data with
  excessive zeros: A case study of head-on heavy vehicle crashes in
  Queensland}. {\it Accident Analysis {\&} Prevention}, 160, 106308.
  doi:10.1016/J.AAP.2021.106308.

\bibitem[{Calcagno and de~Mazancourt(2010)}]{Calcagno2010Glmulti:Models}
Calcagno, V. \& de~Mazancourt, C. (2010) {glmulti: An R Package for Easy
  Automated Model Selection with (Generalized) Linear Models}. {\it Journal of
  Statistical Software}, 34(12), 1--29. doi:10.18637/JSS.V034.I12.
\newline\urlprefix\url{https://www.jstatsoft.org/index.php/jss/article/view/v034i12}

\bibitem[{Darriba et~al.(2012)Darriba, Taboada, Doallo, and
  Posada}]{Darriba2012JModelTestComputing}
Darriba, D., Taboada, G.L., Doallo, R. \& Posada, D. (2012) {jModelTest 2: more
  models, new heuristics and high-performance computing}. {\it Nature methods},
  9(8), 772. doi:10.1038/NMETH.2109.
\newline\urlprefix\url{/pmc/articles/PMC4594756/
  https://www.ncbi.nlm.nih.gov/pmc/articles/PMC4594756/}

\bibitem[{Deb et~al.(2002)Deb, Pratap, Agarwal, and
  Meyarivan}]{Deb2002ANSGA-II}
Deb, K., Pratap, A., Agarwal, S. \& Meyarivan, T. (2002) {A fast and elitist
  multiobjective genetic algorithm: NSGA-II}. {\it IEEE Transactions on
  Evolutionary Computation}, 6(2), 182--197. doi:10.1109/4235.996017.

\bibitem[{Faraway(2016)}]{Faraway2016ExtendingEdition}
Faraway, J.J. (2016) {Extending the Linear Model with R: Generalized Linear,
  Mixed Effects and Nonparametric Regression Models, Second Edition}. {\it
  Extending the Linear Model with R: Generalized Linear, Mixed Effects and
  Nonparametric Regression Models, Second Edition},,
  1--395.doi:10.1201/9781315382722.

\bibitem[{Fiske and Chandler(2011)}]{Fiske2011Unmarked:Abundance}
Fiske, I.J. \& Chandler, R.B. (2011) {unmarked: An R Package for Fitting
  Hierarchical Models of Wildlife Occurrence and Abundance}. {\it Journal of
  Statistical Software}, 43(10), 1--23. doi:10.18637/JSS.V043.I10.
\newline\urlprefix\url{https://www.jstatsoft.org/index.php/jss/article/view/v043i10}

\bibitem[{{Florian Heraud} et~al.(2025){Florian Heraud}, {Behrang Assemi},
  {Douglas Baker}, {Krishna Behara}, {Zeke Ahern}, and {Alexander
  Paz}}]{FlorianHeraud2025UncoveringBehavior}
{Florian Heraud}, {Behrang Assemi}, {Douglas Baker}, {Krishna Behara}, {Zeke
  Ahern} \& {Alexander Paz} (2025) {Uncovering Divergences in Parking Payment
  Behavior}. {\it Case Studies on Transport Policy},.

\bibitem[{Greene(2007)}]{Greene2007FunctionalData}
Greene, W. (2007) {Functional Form and Heterogeneity in Models for Count Data}.
  {\it Foundations and Trends R in Econometrics}, 1(2), 113--218.
  doi:10.1561/0800000008.

\bibitem[{Hilbe(2006)}]{Hilbe2006A4.0}
Hilbe, J.M. (2006) {A review of LIMDEP 9.0 and NLOGIT 4.0}. {\it American
  Statistician}, 60(2), 187--202. doi:10.1198/000313006X110492.

\bibitem[{Hou et~al.(2018)Hou, Tarko, and Meng}]{Hou2018AnalyzingApproach}
Hou, Q., Tarko, A.P. \& Meng, X. (2018) {Analyzing crash frequency in freeway
  tunnels: A correlated random parameters approach}. {\it Accident Analysis
  {\&} Prevention}, 111, 94--100.
  doi:https://doi.org/10.1016/j.aap.2017.11.018.
\newline\urlprefix\url{http://www.sciencedirect.com/science/article/pii/S0001457517304050}

\bibitem[{Hunter et~al.(2023)Hunter, Perry, Salehi, Bandurski, Hubbard, Ball,
  and Morad~Hameed}]{Hunter2023ScienceCare}
Hunter, O.F., Perry, F., Salehi, M., Bandurski, H., Hubbard, A., Ball, C.G.
  et~al. (2023) {Science fiction or clinical reality: a review of the
  applications of artificial intelligence along the continuum of trauma care}.
  {\it World Journal of Emergency Surgery}, 18(1), 16.
  doi:10.1186/s13017-022-00469-1.

\bibitem[{Huo et~al.(2020)Huo, Leng, Hou, and Yang}]{Huo2020AAnalysis}
Huo, X., Leng, J., Hou, Q. \& Yang, H. (2020) {A Correlated Random Parameters
  Model with Heterogeneity in Means to Account for Unobserved Heterogeneity in
  Crash Frequency Analysis}. {\it Transportation Research Record}, 2674(7),
  312--322. doi:10.1177/0361198120922212.
\newline\urlprefix\url{https://journals.sagepub.com/doi/abs/10.1177/0361198120922212}

\bibitem[{Kattan et~al.(2010)Kattan, Abdullah, and
  Salam}]{Kattan2010HarmonyNetworks}
Kattan, A., Abdullah, R. \& Salam, R.A. (2010) {\it {Harmony Search Based
  Supervised Training of Artificial Neural Networks}}.
\newline\urlprefix\url{http://dx.doi.org/10.1109/isms.2010.31}

\bibitem[{Kim et~al.(2023)Kim, Holton, Sweeting, Koreshe, McGeechan, and
  Miskovic-Wheatley}]{Kim2023UsingDisorders}
Kim, M., Holton, M., Sweeting, A., Koreshe, E., McGeechan, K. \&
  Miskovic-Wheatley, J. (2023) {Using health administrative data to model
  associations and predict hospital admissions and length of stay for people
  with eating disorders}. {\it BMC psychiatry}, 23(1), 326.
  doi:10.1186/S12888-023-04688-X/TABLES/6.
\newline\urlprefix\url{https://bmcpsychiatry.biomedcentral.com/articles/10.1186/s12888-023-04688-x}

\bibitem[{Lefort et~al.(2017)Lefort, Longueville, and
  Gascuel}]{Lefort2017SMS:PhyML}
Lefort, V., Longueville, J.E. \& Gascuel, O. (2017) {SMS: Smart Model Selection
  in PhyML}. {\it Molecular Biology and Evolution}, 34(9), 2422--2424.
  doi:10.1093/MOLBEV/MSX149.
\newline\urlprefix\url{https://dx.doi.org/10.1093/molbev/msx149}

\bibitem[{Li et~al.(2007)Li, Harman, and Hierons}]{Li2007SearchPrioritization}
Li, Z., Harman, M. \& Hierons, R.M. (2007) {Search algorithms for regression
  test case prioritization}. {\it IEEE Transactions on Software Engineering},
  33(4), 225--237. doi:10.1109/TSE.2007.38.

\bibitem[{Lord and Mannering(2010)}]{Lord2010TheAlternatives}
Lord, D. \& Mannering, F. (2010) {The statistical analysis of crash-frequency
  data: A review and assessment of methodological alternatives}. {\it
  Transportation Research Part A: Policy and Practice}, 44(5), 291--305.
  doi:10.1016/j.tra.2010.02.001.

\bibitem[{Lord et~al.(2005)Lord, Washington, and Ivan}]{Lord2005PoissonTheory}
Lord, D., Washington, S.P. \& Ivan, J.N. (2005) {Poisson, Poisson-gamma and
  zero-inflated regression models of motor vehicle crashes: balancing
  statistical fit and theory}. {\it Accident Analysis {\&} Prevention}, 37(1),
  35--46. doi:https://doi.org/10.1016/j.aap.2004.02.004.
\newline\urlprefix\url{http://www.sciencedirect.com/science/article/pii/S0001457504000521}

\bibitem[{Mannering et~al.(2020)Mannering, Bhat, Shankar, and
  Abdel-Aty}]{Mannering2020BigAnalysis}
Mannering, F., Bhat, C.R., Shankar, V. \& Abdel-Aty, M. (2020) {Big data,
  traditional data and the tradeoffs between prediction and causality in
  highway-safety analysis}. {\it Analytic methods in accident research}, 25,
  100113.

\bibitem[{Mannering and Bhat(2014)}]{Mannering2014AnalyticDirections}
Mannering, F.L. \& Bhat, C.R. (2014) {Analytic methods in accident research:
  Methodological frontier and future directions}. {\it Analytic Methods in
  Accident Research}, 1, 1--22. doi:10.1016/j.amar.2013.09.001.

\bibitem[{Marchal et~al.(2017)Marchal, Cumming, and
  McIntire}]{Marchal2017ExploitingCanada}
Marchal, J., Cumming, S.G. \& McIntire, E.J. (2017) {Exploiting Poisson
  additivity to predict fire frequency from maps of fire weather and land cover
  in boreal forests of Qu{\'{e}}bec, Canada}. {\it Ecography}, 40(1), 200--209.
  doi:10.1111/ECOG.01849.

\bibitem[{Miller(2002)}]{Miller2002SubsetRegression}
Miller, A. (2002) {\it {Subset Selection in Regression}}. : Chapman and
  Hall/CRC.
\newline\urlprefix\url{https://www.taylorfrancis.com/books/mono/10.1201/9781420035933/subset-selection-regression-alan-miller}

\bibitem[{Milton and Mannering(1998)}]{Milton1998TheFrequencies}
Milton, J. \& Mannering, F. (1998) {The relationship among highway geometrics,
  traffic-related elements and motor-vehicle accident frequencies}. {\it
  Transportation}, 25(4), 395--413. doi:10.1023/A:1005095725001/METRICS.
\newline\urlprefix\url{https://link.springer.com/article/10.1023/A:1005095725001}

\bibitem[{Oliveira et~al.(2010)Oliveira, Braga, Lima, and
  Corn{\'{e}}lio}]{Oliveira2010GA-basedEstimation}
Oliveira, A.L., Braga, P.L., Lima, R.M. \& Corn{\'{e}}lio, M.L. (2010)
  {GA-based method for feature selection and parameters optimization for
  machine learning regression applied to software effort estimation}. {\it
  Information and Software Technology}, 52(11), 1155--1166.
  doi:10.1016/J.INFSOF.2010.05.009.
\newline\urlprefix\url{https://www.sciencedirect.com/science/article/abs/pii/S0950584910000984}

\bibitem[{Paz et~al.(2019)Paz, Arteaga, and
  Cobos}]{Paz2019SpecificationFramework}
Paz, A., Arteaga, C. \& Cobos, C. (2019) {Specification of mixed logit models
  assisted by an optimization framework}. {\it Journal of Choice Modelling},
  30, 50--60. doi:10.1016/j.jocm.2019.01.001.
\newline\urlprefix\url{http://dx.doi.org/10.1016/j.jocm.2019.01.001}

\bibitem[{Sadeghkhani(2022)}]{Sadeghkhani2022K1K2InflatedMatches}
Sadeghkhani, A. (2022) {K1K2–Inflated Conway–Maxwell–Poisson Model:
  Bayesian Predictive Modeling with an Application in Soccer Matches}. {\it
  American Journal of Mathematical and Management Sciences}, 41(4), 295--304.
  doi:10.1080/01966324.2021.1960225.
\newline\urlprefix\url{https://www.tandfonline.com/doi/abs/10.1080/01966324.2021.1960225}

\bibitem[{Saeed et~al.(2019)Saeed, Hall, Baroud, and
  Volovski}]{Saeed2019AnalyzingHighways}
Saeed, T.U., Hall, T., Baroud, H. \& Volovski, M.J. (2019) {Analyzing road
  crash frequencies with uncorrelated and correlated random-parameters count
  models: An empirical assessment of multilane highways}. {\it Analytic methods
  in accident research}, 23, 100101.

\bibitem[{Sawtelle et~al.(2023)Sawtelle, Shirazi, Garder, and
  Rubin}]{Sawtelle2023DriverMaine}
Sawtelle, A., Shirazi, M., Garder, P.E. \& Rubin, J. (2023) {Driver, roadway,
  and weather factors on severity of lane departure crashes in Maine}. {\it
  Journal of Safety Research}, 84, 306--315. doi:10.1016/J.JSR.2022.11.006.

\bibitem[{Sellers and Shmueli(2010)}]{Sellers2010AData}
Sellers, K.F. \& Shmueli, G. (2010) {A flexible regression model for count
  data}. {\it Annals of Applied Statistics}, 4(2), 943--961.
  doi:10.1214/09-AOAS306.

\bibitem[{Shankar et~al.(1996)Shankar, Mannering, and
  Barfield}]{Shankar1996StatisticalFreeways}
Shankar, V., Mannering, F. \& Barfield, W. (1996) {Statistical analysis of
  accident severity on rural freeways}. {\it Accid. Anal. and Prev}, 28(3).

\bibitem[{Stasinopoulos and Rigby(2008)}]{Stasinopoulos2008GeneralizedR}
Stasinopoulos, D.M. \& Rigby, R.A. (2008) {Generalized Additive Models for
  Location Scale and Shape (GAMLSS) in R}. {\it Journal of Statistical
  Software}, 23(7), 1--46. doi:10.18637/JSS.V023.I07.
\newline\urlprefix\url{https://www.jstatsoft.org/index.php/jss/article/view/v023i07}

\bibitem[{Train(2009)}]{Train2009DiscreteSimulation}
Train, K.E. (2009) {\it {Discrete Choice Methods with Simulation}}. No.
  9780521766555 in Cambridge Books. : Cambridge University Press.
\newline\urlprefix\url{https://ideas.repec.org/b/cup/cbooks/9780521766555.html}

\bibitem[{Venables and Ripley(1999)}]{Venables1999ModernS-PLUS}
Venables, W.N. \& Ripley, B.D. (1999) {\it {Modern Applied Statistics with
  S-PLUS}}. : American Statistical Association.

\bibitem[{Wu et~al.(2018)Wu, Shen, Li, Chen, Lin, and
  Suganthan}]{Wu2018EnsembleVariants}
Wu, G., Shen, X., Li, H., Chen, H., Lin, A. \& Suganthan, P.N. (2018) {Ensemble
  of differential evolution variants}. {\it Information Sciences}, 423,
  172--186. doi:10.1016/j.ins.2017.09.053.

\bibitem[{Yee(2010)}]{Yee2010TheAnalysis}
Yee, T.W. (2010) {The VGAM Package for Categorical Data Analysis}. {\it Journal
  of Statistical Software}, 32(10), 1--34. doi:10.18637/JSS.V032.I10.
\newline\urlprefix\url{https://www.jstatsoft.org/index.php/jss/article/view/v032i10}

\bibitem[{Zeileis et~al.(2008)Zeileis, Kleiber, and
  Jackman}]{Zeileis2008RegressionR}
Zeileis, A., Kleiber, C. \& Jackman, S. (2008) {Regression Models for Count
  Data in R}. {\it Journal of Statistical Software}, 27(8), 1--25.
  doi:10.18637/JSS.V027.I08.
\newline\urlprefix\url{https://www.jstatsoft.org/index.php/jss/article/view/v027i08}

\end{thebibliography}

\section*{Supporting information}
Package can be installed here:  \url{https://pypi.org/project/metacountregressor/}
Notebooks Here: \url{https://anonymous.4open.science/r/data-8257}

\section{Key Arguments for the software}
\onecolumn

\begin{longtable}
{|K{0.32\textwidth}|K{0.62\textwidth}|}
\caption{Description of Solution Arguments and Parameters} \label{tab:definedargs} \\ \hline
\rowcolor{red}\textit{Argument} & \textit{Description} \\ \hline
\rowcolor{yellow}\multicolumn{2}{|l|}{\textbf{Solution Arguments: (entered as \code{*args} into the function \code{ObjectiveFunction})}} \\ \hline
\endfirsthead

\rowcolor{white}\multicolumn{2}{|l|}{\ldots continued} \\ \hline
\textit{Argument} & \textit{Description} \\ \hline
\endhead

\hline
\rowcolor{white}\multicolumn{2}{|r@{}}{continued \ldots} \\ \hline
\endfoot
\endlastfoot

-\_obj\_1 = 'BIC' & Information criterion to minimize. Available options are "BIC", "AIC", "HQIC", "CAIC", and "AICc". \\ \hline
-complexity\_level = 6 & Integers or a list of integers that encapsulate the hierarchical capabilities of the proposed framework. 5 represents heterogeneity in the means, 4 represents grouped random parameters, 3 represents correlated random parameters, 2 represents standard random parameters, 1 represents fixed effects, and 0 represents off. Entering the integer "5" will include specifications of all reduced levels. To exclude specific tests, provide a list of elements (e.g., [1,2,3,4,5] excludes grouped random parameters). \\ \hline
-model\_types = [[0,1]] & List of model types. 0 for Poisson, 1 for Negative Binomial. \\ \hline
-\_distributions = ['t', 'u', 'n', 'ln\_n', 'tn\_n'] & List of available distributions for the random parameters. Includes triangular, uniform, normal, log-normal, and truncated normal. \\ \hline
-\_transformations = ['no', 'sqrt', 'log', 'arcsinh'] & List of possible transformations to define. Others that can be added to the list are 'exp', 'fact'. \\ \hline
-\_max\_time=3600 & Maximum amount of CPU time before termination. \\ \hline
-method\_ll = 'L-BFGS-B' & Selected optimization algorithm for solving the lower-level likelihood. Choose one from 'L-BFGS-B', 'BFGS\_2', or 'Nelder-Mead-BFGS'. \\ \hline
-\_max\_characteristics = 25 & Integer number to preclude the potential number of contributing factors to consider in the model. \\ \hline
\rowcolor{yellow}\multicolumn{2}{|l|}{\textbf{Global Algorithmic Parameters for All Metaheuristic Algorithms:}} \\ \hline
-m=1000, --\_max\_iter=1000 & Maximum number of iterations. \\ \hline
\rowcolor{yellow}\multicolumn{2}{|l|}{\textbf{DE Specific Parameters:}} \\ \hline
-\_AI=1 & Adjustment index for transitioning into new decision states, which are ordered. The index value represents the allowable change in the decision spaces. \\ \hline
-\_cr=0.2 & Decimal value less than 1.00 such that it represents the percentage chance of applying a crossover on a chromosome element. \\ \hline
-\_pop\_size = 20 & Population size for the algorithm. \\ \hline
\rowcolor{yellow}\multicolumn{2}{|l|}{\textbf{SA Specific Parameters:}} \\ \hline
-m=15, --max\_iter=15 & Maximum number of iterations. \\ \hline
-alpha = 0.95 & Temperature reduction factor. \\ \hline
-\_ts=2 & Number of iterations before the temperature is reduced again by -alpha. \\ \hline
\rowcolor{yellow}\multicolumn{2}{|l|}{\textbf{HS Specific Parameters:}} \\ \hline
-\_hms=20, --\_pop\_size = 20 & Harmony memory size for the algorithm. \\ \hline
-\_par=0.4 & Pitch adjustment rate. \\ \hline
-\_hmcr = 0.4 & Harmony memory consideration rate. \\ \hline
-\_AI=1 & Adjustment index for transitioning into new-state spaces. \\ \hline
\end{longtable}

\subsection{Glossary \label{app1.1a}}
\printglossaries

\end{document}